\documentclass[aps,prb,twocolumn,superscriptaddress,amsmath]{revtex4-2}

\usepackage{graphicx}
\usepackage{hyperref}
\usepackage{mathrsfs}
\usepackage{bm}
\usepackage{color}
\usepackage{siunitx}
\usepackage[capitalize]{cleveref}
\hypersetup{hypertex=true,
	  colorlinks=true,
	anchorcolor=blue,
	linkcolor=blue,
        citecolor=blue,
	urlcolor=blue}

\usepackage[version=3]{mhchem}
\usepackage{appendix}
\usepackage{amssymb}
\usepackage{xcolor}
\usepackage{url}

\begin{document}

\preprint{APS/123-QED}

\title{Unconventional Spin Valve Based on Normal Metal/Chiral Molecule/Altermagnet Junctions}

\author{Tian-Yi Zhang}
\affiliation{International Center for Quantum Materials, School of Physics, Peking University, Beijing 100871, China}

\author{Peng-Yi Liu}
\affiliation{International Center for Quantum Materials, School of Physics, Peking University, Beijing 100871, China}

\author{Yu-Fei Sun}
\affiliation{International Center for Quantum Materials, School of Physics, Peking University, Beijing 100871, China}
\affiliation{Hefei National Laboratory, Hefei 230088, China}

\author{Ai-Min Guo}
\affiliation{Hunan Key Laboratory for Super-microstructure and Ultrafast Process, School of Physics, Central South University, Changsha 410083, China}

\author{Qing-Feng Sun}
\email{sunqf@pku.edu.cn}
\affiliation{International Center for Quantum Materials, School of Physics, Peking University, Beijing 100871, China}
\affiliation{Hefei National Laboratory, Hefei 230088, China}

\begin{abstract}
Chiral molecules have attracted broad interdisciplinary interest for their ability to produce highly spin-polarized current.
This phenomenon, known as the chiral-induced spin selectivity effect, holds great potential in the field of spintronics.
Here, we propose to combine chiral molecules with altermagnets to construct highly efficient and tunable spin valves.
Using the nonequilibrium Green's function method and the Landauer-Büttiker formula, we obtain the conductance and the magnetoresistance of a normal metal/chiral molecule/altermagnet spin valve.
Our theoretical results reveal that the conductance of the spin valve can be effectively tuned by reorienting the Néel vector of the altermagnet, and the magnetoresistance of the spin valve increases with molecular length and altermagnetic anisotropy.
Moreover, the magnetoresistance vanishes for achiral molecules or in the absence of molecular spin-orbit coupling.
Our work paves the way for developing efficient, controllable, and stray-field-free spintronic devices.
\end{abstract}

\maketitle

\section{\label{SEC1} Introduction}

Chirality and chiral substances are ubiquitous in nature.
The unique properties arising from this structural asymmetry have attracted broad research interest across physics, chemistry, biology, and materials science \cite{Hsieh2009,Bornscheuer2012}.
A landmark discovery in this field was made by Naaman \textit{et al.}, who demonstrated that unpolarized electrons become spin-polarized after transmitting through chiral, nonmagnetic stearoyl lysine molecules \cite{Ray1999}.
This phenomenon, known as the chiral-induced spin selectivity (CISS) effect \cite{Xu2023,Bloom2024,Foo2025,LLi2025}, reveals a profound connection between molecular chirality and another fundamental property of nature: electron spin.
Since its discovery, the CISS effect has been observed across a broad range of chiral materials \cite{Gohler2011,Sun2024,Wang2024,Moharana2025,Mishra2013,Kiran2016,Jia2021,Abendroth2019,Levi2016,Kumar2017,Eckvahl2023,Eckvahl2024,Latawiec2025,Eckvahl2026,TLiu2024}.
Experiments consistently show that chiral molecules can generate spin-polarized currents, with the spin polarization direction reversing upon inversion of the material chirality \cite{Xu2023,Bloom2024,Sun2024,Kiran2016}.
To explain the CISS effect, various mechanisms have been proposed \cite{Guo2012,Guo2014,Chen2024,Alwan2021,Chiesa2024,Fransson2023,Dianat2020,Sierra2020,Wang2026,Kundu2024,Sun2026}.
One effective mechanism suggests that the combination of molecular chirality with spin-orbit coupling (SOC) and dephasing induces the CISS effect \cite{Guo2012,Guo2014,LLiu2026}.
This mechanism provides a unified explanation for longitudinal and transverse CISS effects \cite{Guo2012,Guo2014,Wang2024}, the inverse CISS effect \cite{Zhang2025c}, the dynamic processes of CISS \cite{Zhang2025a,Liu2025a}, the giant magnetoresistance (MR) phenomenon in chiral molecule/ferromagnet (FM) junctions \cite{Zhang2025b}, and more \cite{Liu2025b}. Currently, the field of CISS is undergoing rapid development and is attracting increasing attention.

Recently, the CISS effect has been proposed to construct a new type of spin valve \cite{Adhikari2023,Bloom2024,Bloom2025}.
As a fundamental building block in spintronics, a conventional spin valve consists of a fixed FM layer and a free FM layer separated by a normal metal (NM) \cite{Julliere1975,Moodera1995}.
Its conductance depends on the relative magnetization alignment of the two FMs, resulting in an MR effect due to spin-dependent electron transport \cite{Julliere1975,Moodera1995,SunYF2025}.
However, such a spin valve faces several inherent limitations: low FM spin polarization, susceptibility of the fixed layer to change under external fields, undesirable stray magnetic fields that cause crosstalk, and low FM precession frequencies (around GHz) \cite{Bai2024}.
To overcome these, it has been proposed to replace the fixed FM layer with chiral molecules that exhibit the CISS effect \cite{Adhikari2023}.
CISS provides highly spin-polarized currents that are unaffected by external fields, thereby mitigating the issues of low polarization and unwanted field interference \cite{Xu2023,Bloom2025}.
Nevertheless, since the free layer remains FM, challenges related to stray fields and switching speed persist, presenting a key bottleneck for further development \cite{Bloom2024}.

The emergence of altermagnet (AM) offers a breakthrough pathway to overcome this bottleneck \cite{Smejkal2022a,Smejkal2022b,Smejkal2022c,Amin2024,Yamamoto2025}.
AM is a novel magnet phase characterized by a compensated antiparallel magnetic order in real space, where opposing spin sublattices are linked by crystal rotation symmetries \cite{Cheng2024,Bai2024,Zhou2025,Yi2025,SunYF2025}.
Despite the absence of macroscopic magnetization, it still breaks time-reversal symmetry and induces a strong momentum-dependent spin splitting, with the splitting energy reaching the order of 1 eV \cite{Jiang2025,ZhangX2025,Bai2024}.
Experimentally, the magnetic order (Néel vector) of AM can be controlled through electrical and magnetic methods \cite{Han2024,Zhou2025,Bai2024,YZhang2025,Li2025,Leiviska2024,Rial2024}, and a variety of intriguing spin-dependent transport phenomena have been observed \cite{Zhu2025,Bai2024}.
Importantly, the momentum-dependent spin splitting enables AMs to efficiently probe spin-polarized currents, much like FMs \cite{Smejkal2022a,Smejkal2022b,Bai2024}.
With these fascinating properties, an attractive prospect is to integrate AMs with chiral molecules to construct a new type of spin valve.
Owing to the compensated magnetic order without stray fields and the THz-scale magnetic dynamics of AM, such a chiral molecule/AM spin valve is expected to overcome all key limitations of conventional spin valves based on FMs.

\begin{figure}[t]
\centering
\includegraphics[width = 1.0 \linewidth]{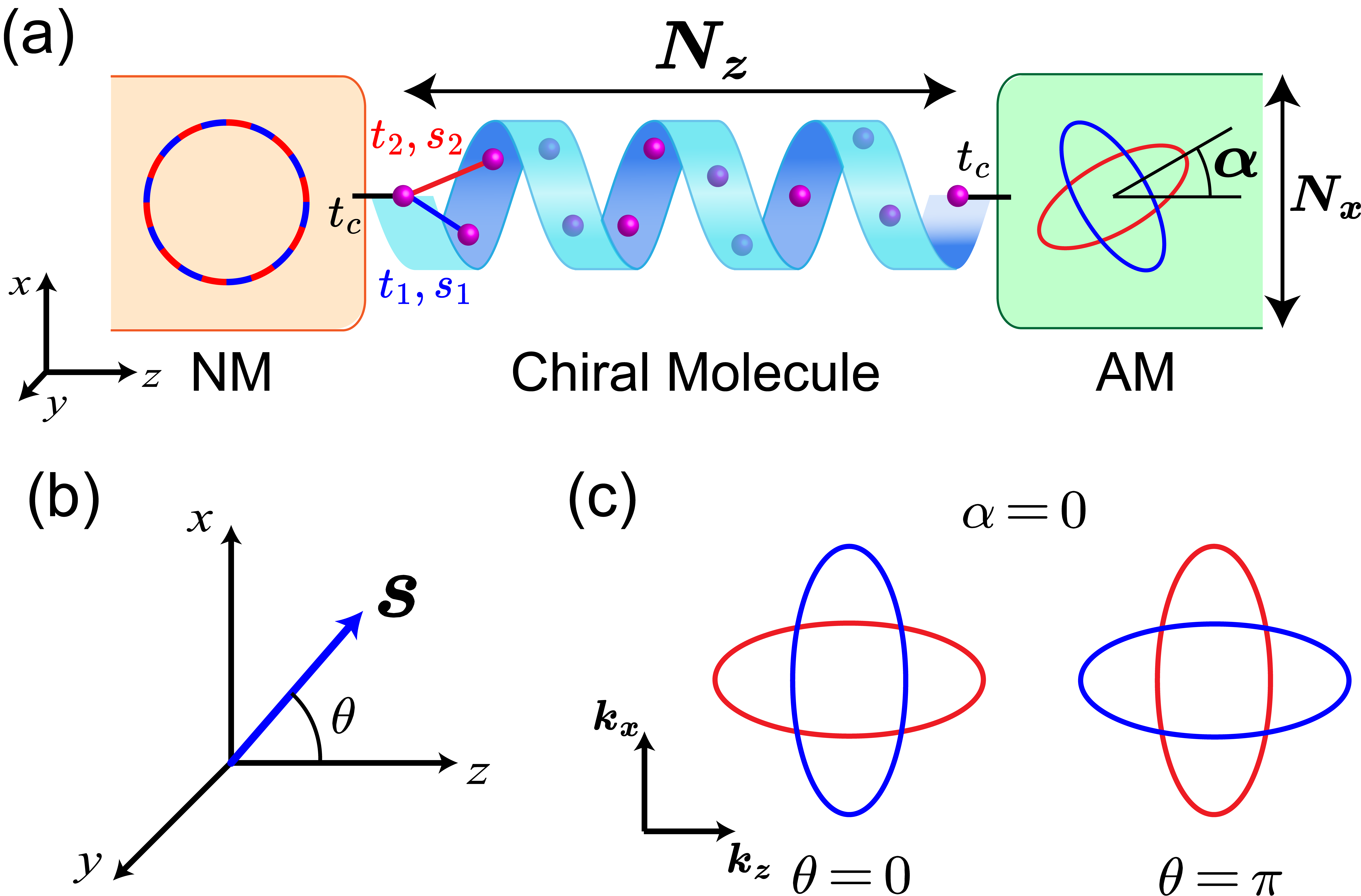}
\caption{\label{fig1}
(a) Schematic of the device:
The NM (orange region) and AM (green region) electrodes are placed on the left and right sides of the chiral molecule, respectively.
Spin-up and spin-down Fermi surfaces are marked in blue and red, respectively.
$\alpha$ is the angle between the crystalline axis and the $+z$ axis, denoting the orientation of the AM.
The widths of the electrodes are $N_x$, and the length of the molecule is $N_z$.
$t_j, \ s_j$ ($j=1,2$) are hopping integral and SOC, and $t_c$ is the coupling energy of chiral molecule and electrodes.
(b) The orientation of the Néel vector $\bm{s}$ for the AM electrode.
It is oriented within the $xz$ plane, defined by the angle $\theta$.
(c) The Fermi surfaces of AM for $\theta=0$ (left panel) and $\theta=\pi$ (right panel) with $\alpha=0$.}
\end{figure}

In this paper, we theoretically investigate a system with a chiral molecule sandwiched between an NM and an AM electrode, forming an NM/chiral molecule/AM junction, as illustrated in Fig. \ref{fig1}(a).
We show that the conductance of this junction depends on the orientation of the AM's Néel vector and is also influenced by the crystallographic orientation of the AM.
We demonstrate that the MR of this system arises from the interplay of the chirality, molecular SOC, and the momentum-dependent spin splitting of the AM.
These contributions are collectively necessary, and the MR increases with both the length of the molecule and the degree of AM anisotropy.
This work paves a new pathway for the development of spintronic devices.

The rest of the paper is organized as follows. In Sec.~\ref{SEC2}, we introduce the model Hamiltonian of the spin valve.
In Sec.~\ref{SEC3}, we study the transport properties of the spin valve.
In Sec.~\ref{SEC4}, we analyze the current and the MR of the spin valve at finite bias.
Finally, we provide a discussion and a summary in Sec.~\ref{SEC5}.

\section{\label{SEC2} Model and Hamiltonian}

We begin by introducing the NM/chiral molecule/AM junction model.
As shown in Fig. \ref{fig1}(a), the left side in orange is the semi-infinite NM electrode extending along the $-z$ direction, and the right side in green is the semi-infinite AM electrode extending along the $+z$ direction.
Both electrodes have a finite width $N_x$ along the $x$ direction.
In the middle, a chiral molecule with length $N_z$ is sandwiched between the two electrodes.
When a voltage is applied between the two electrodes, a current will flow along the $z$ direction across the molecule.
The whole system can be simulated by the Hamiltonian: $H=H_{AM}+H_{NM}+H_{cmol}+H_{T}+H_d$.
Here, the first and second terms $H_{AM}$, $H_{NM}$ describe the AM and NM electrodes, respectively.
The third term $H_{cmol}$ describes the chiral molecule \cite{Guo2014}, and $H_{T}$ describes the coupling.
$H_d$ describes the dephasing in the chiral molecule \cite{Guo2012}.
In the following, we explain each term in detail.

In the continuum model, the Hamiltonian of the AM is: $H_{AM}(k_x,k_z)=t_0(k_x^2+k_z^2) + t_J [(k_z^2-k_x^2)\cos 2 \alpha + 2 k_x k_z \sin 2 \alpha] \bm{\sigma} \cdot \bm{s}$.
Here, $t_0$ is the kinetic coefficient, and $t_J$ is the magnitude of AM.
As shown in Fig. \ref{fig1}(a), in the AM electrode, the angle $\alpha$ between the crystalline axis and the $+z$ axis defines the orientation of AM.
$\bm{\sigma}=(\sigma_x,\sigma_y,\sigma_z)$ are the Pauli matrices.
$\bm{s}$ is the Néel vector that determines the orientation of spins on the Fermi surface, as shown in Fig. \ref{fig1}(b).
In this system, transport is governed by the angle between $\bm{s}$ and spin polarization direction of the molecule \cite{SunYF2025}.
Since the molecular spin polarization is primarily along the $z$ direction, we can, without loss of generality, constrain $\bm{s}$ to vary within the $xz$ plane.
Therefore, $\bm{s}=(\sin \theta, 0, \cos \theta)$ is uniquely determined by parameter $\theta$.
The Fermi surface of two spins consists of two ellipses with mutually perpendicular major axes \cite{SunYF2025}.
In Fig. \ref{fig1}(a), $\theta = 0$, and the blue and red ellipses represent spin-up and spin-down, respectively.
In real experiments, the spin direction on the Fermi surface can be altered by tuning the direction of the Néel vector \cite{Han2024,Zhou2025}.
Figure \ref{fig1}(c) displays the Fermi surfaces for different spins in the AM at $\alpha = 0$, under two different Néel vector directions $\theta=0,\pi$.
When $\theta = 0$, the major axis of the spin-up and spin-down Fermi surfaces lies along the $k_x$ and $k_z$ directions, respectively, resulting in anisotropies in the Fermi surfaces for different spins.
When $\theta = \pi$, the situation is exactly the opposite.
In the NM electrode, $H_{NM}(k_x,k_z)=t_0(k_x^2+k_z^2)$, the Fermi surfaces for the two spin species are two completely overlapping circles.
To calculate the conductance of the junction, we discretize $H_{AM}, H_{NM}$ on a two-dimensional square lattice, shown in Appendix~\ref{secA}.

As shown in Fig. \ref{fig1}(a), the molecule is single-helical, and can be described by the Hamiltonian \cite{Guo2014,Zhang2025b}:
\begin{equation}
    \begin{aligned}
    H_{cmol}=&-\mu\sum_{n_z=1}^{N_z} c_{n_z}^{\dagger}c_{n_z}+\sum_{n_z=1}^{N_z-1} \sum_{j=1}^{N_z-n_z}\left(t_j c_{n_z}^\dagger c_{n_z+j}+ \text{H.c.}\right) \\
    & + \sum_{n_z=1}^{N_z-1} \sum_{j=1}^{N_z-n_z}\left[2\text{i}s_j \cos(\varphi_{n_z,j}^{-}) c_{n_z}^\dagger \sigma_{n_z,j} c_{n_z+j}+\text{H.c.}\right],
    \end{aligned}
\label{Hcmol}
\end{equation}
where $c_{n_z}^{\dagger}=(c_{n_z,\uparrow}^{\dagger},c_{n_z,\downarrow}^{\dagger})$ is the creation operator at the site $n_z$.
The length of the chiral molecule is $N_z$.
$\mu$ is the chemical potential.
$t_j=t_1 e^{-(l_j-l_1)/l_c}$ and $s_j=s_1 e^{-(l_j-l_1)/l_c}$ are hopping integral between sites $n_z$ and $n_z+j$ and the corresponding SOC, respectively. 
$l_c$ is the decay exponent.
$l_j=\sqrt{[2R \sin (\Delta \varphi /2)]^2+(j\Delta h)^2}$ is the distance between sites $n_z$ and $n_z+j$, with $R$ being the molecular radius. 
$\Delta \varphi$ and $\Delta h$ are the twist angle and the stacking distance between two neighboring sites along the molecular axis, respectively.
$\sigma_{n_z,j}=\left[\sigma_x \sin (\varphi_{n_z,j}^{+})-\sigma_y \cos (\varphi_{n_z,j}^{+}) \right] \sin \theta_j + \sigma_z \cos \theta_j$, $\varphi_{n_z,j}^{\pm}=(\varphi_{n_z+j} \pm \varphi_{n_z})/2$, and $\varphi_{n_z}=n_z \Delta \varphi$ is the cylindrical coordinate of the site $n_z$. 
$\theta_j=\arccos[2R \sin (j \Delta \varphi/2)/l_j]$ is the space angle.
Further discussion of the model is provided in Appendix~\ref{secA}.

In Fig. \ref{fig1}(a) $t_1, s_1, t_2$ and $s_2$ are schematically illustrated as examples.
This model is based on our previous work, which provides a consistent framework for understanding diverse CISS-related phenomena \cite{Liu2025a,Liu2025b,Zhang2025b,Zhang2025c}.
The chiral molecule is connected at the middle of the electrodes with coupling energy $t_c$ (see Appendix~\ref{secA} for the details).
In real systems, impurities, electron-electron interaction, and electron-phonon interaction inevitably cause the dephasing effect, which is shown to play an important role in molecules \cite{Morita2003,Xing2008,Skourtis2005,Zhang2025a}.
To simulate these processes, we introduce the Büttiker's virtual electrodes connecting to each site of the chiral molecule, with the Hamiltonian being $H_d=\sum_{k}\sum_{n_z=1}^{N_z} (\varepsilon_{n_z,k}b_{n_z,k}^{\dagger} b_{n_z,k} + t_d b_{n_z,k}^{\dagger} c_{n_z} + \text{H.c.})$ \cite{Guo2012,Guo2014,Buttiker1986}.
$b_{n_z,k}^{\dagger}=(b_{n_z,k,\uparrow}^{\dagger},b_{n_z,k,\downarrow}^{\dagger})$ is the creation operator of the virtual electrode.
$\varepsilon_{n_z,k}$ is the spectrum of each virtual electrode and $t_d$ describes the coupling.
Based on the above Hamiltonian, we calculate the conductance $G$ of the system \cite{Datta1995,Lee1981}.
The methods are provided in Appendix~\ref{secB}.

For the chiral molecule, the structural parameters are $R=0.25 \ \text{nm}$, $\Delta h = 0.15 \ \text{nm}$.
The twist angle is set to $\Delta \varphi=5\pi/9$ ($-5\pi/9$) for right- (left-) handed molecules, and $\Delta \varphi=0$ for achiral molecules.
The length of the molecule is $N_z=30$, and $l_c=0.9$ \r{A}.
The nearest-neighbor hopping $t_1=100 \ \text{meV}$ is taken as the energy unit.
The SOC is set as $s_1=0.05t_1=5 \ \text{meV}$, which aligns with the SOC strength (a few meV) in chiral molecules composed of light elements \cite{Evers2022,Steele2013,Jhang2010}.
For the AM electrode, we take $t_0=3t_1$ and $t_J=1.5t_1$, which are obtained by fitting the first-principles calculation results and correspond to a spin splitting of about 300 meV \cite{Jiang2025}.
The width of NM and AM electrodes is $N_x=41$, and $t_c=t_1$.

\begin{figure}[t]
\centering
\includegraphics[width = 1.0 \linewidth]{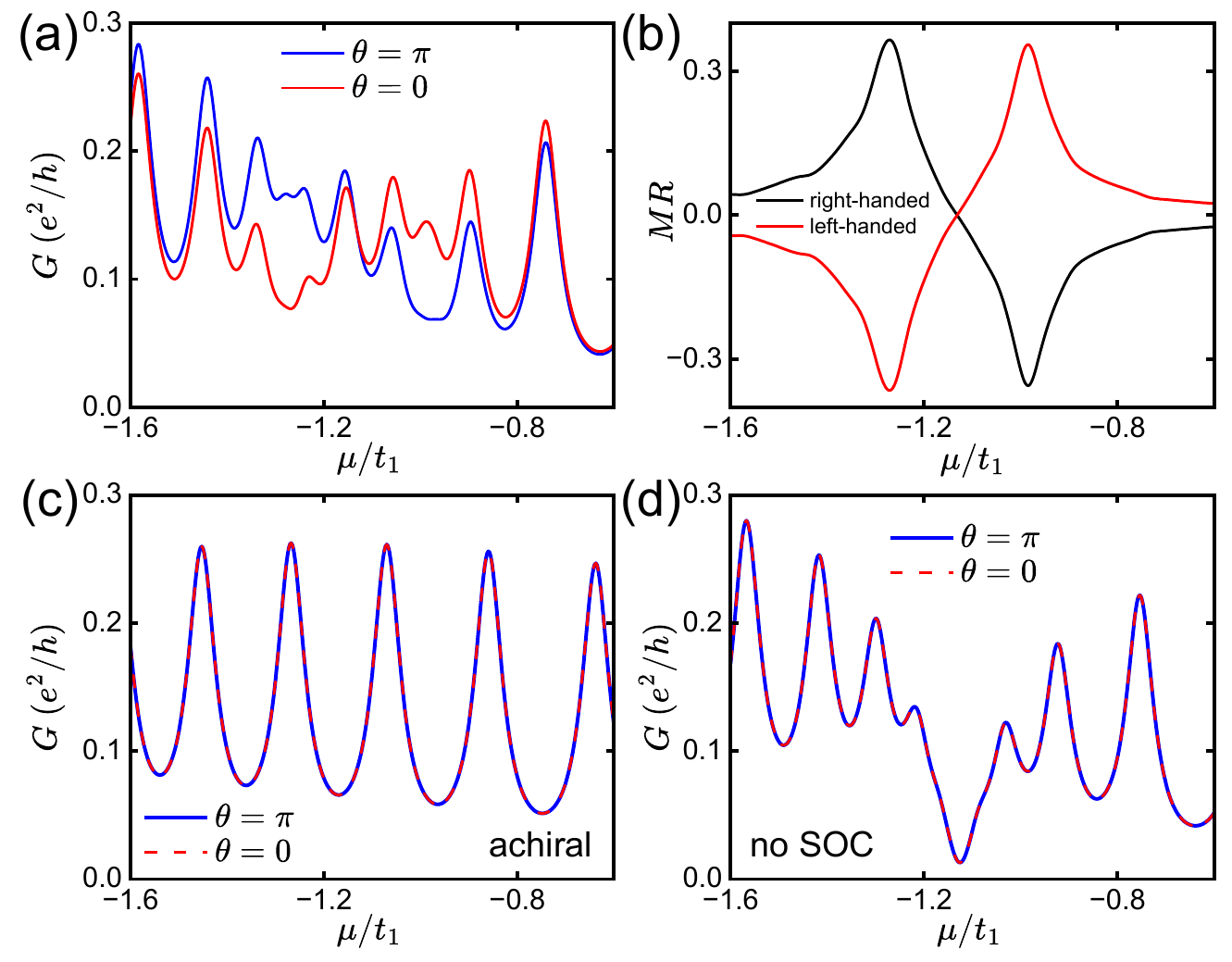}
 \caption{\label{fig2}
 (a) Conductance $G$ versus chemical potential $\mu$ in NM/right-handed molecule/AM device.
 Blue and red lines are the conductances when $\theta=\pi$ and $\theta=0$, respectively.
 (b) MR versus chemical potential.
 Black and red lines are the MR when the molecule is right- and left-handed, respectively.
 (c) Conductance versus chemical potential in the NM/achiral molecule/AM device.
 (d) Conductance versus chemical potential in the NM/no SOC molecule/AM device ($s_1 = 0$).
 The parameters are $E=t_1$ and $\alpha=0$.
 }
\end{figure}

\section{\label{SEC3} Transport behaviors}

Let us study the transport behavior of the system under the opposite orientations of the Néel vector, with the AM orientation angle $\alpha=0$ and Fermi energy $E=t_1$.
Figure \ref{fig2}(a) shows the chemical potential $\mu$ dependence of conductances (blue line for $\theta=\pi$ and red line for $\theta=0$) of a right-handed chiral molecule.
In this conductance spectrum, it clearly shows that $G(\theta=\pi)$ is different from $G(\theta=0)$.
The variation of conductance with the modulation of the AM is similar to that observed in our earlier study on chiral molecule/FM junctions \cite{Zhang2025b}, where the conductance could likewise be tuned by adjusting the FM orientation.
However, the conductance difference here is more intriguing than the FM case, since neither the chiral molecule nor the NM and AM electrodes exhibit macroscopic magnetism.
This difference in conductance between the cases of $\theta=0$ and $\theta=\pi$ arises from the CISS effect in the molecule and the momentum-dependent spin splitting in the AM \cite{Bloom2024,Bai2024}.
The black line in Fig. \ref{fig2}(b) shows the MR derived from Fig. \ref{fig2}(a) using $MR = [G(\theta=\pi) - G(\theta=0)] / [G(\theta=\pi) + G(\theta=0)]$.
In this system, MR can achieve a maximum value of 36.5\% (when $\mu=-1.27t_1$), despite the complete absence of macroscopic magnetism.
The red line in Fig.~\ref{fig2}(b) shows the MR curve for the left-handed molecule, which is exactly opposite to that of the right-handed molecule, a clear signature of the CISS effect \cite{Bloom2024}.

In Appendix~\ref{secB} and Figs.~\ref{figB1}(a,b) therein, we calculate the spin-resolved conductance and spin polarization of the NM/chiral molecule/NM junction, which shows that the current passing through the chiral molecule is indeed spin-polarized.
For comparison, in Fig.~\ref{figB2} in Appendix~\ref{secB}, we calculate the conductance and MR of the NM/chiral molecule/FM junction. 
The results show that compared with the FM case, the AM electrode can achieve similar conductance modulation and does not reduce the efficiency (see Appendix~\ref{secB} for the details).

In the CISS effect, both molecular chirality and SOC play essential roles \cite{Zhang2025b}.
To investigate the significance of these factors in the present system, we show the conductance spectrum in the achiral molecule [Fig. \ref{fig2}(c)] and right-handed molecule without SOC [Fig. \ref{fig2}(d)].
In both cases, $G(\theta=\pi)$ and $G(\theta=0)$ completely overlap, indicating that $MR=0$, although the conductance remains finite in both cases.
In these two cases, the CISS is absent, and the molecule cannot produce spin-polarized current \cite{Zhang2025b}.
Although the AM itself possesses strong spin splitting, this property remains ineffective in generating MR when the incoming current is spin-unpolarized.
Figure~\ref{figB1}(c) in Appendix~\ref{secB} shows the conductance of the NM/chiral molecule/NM system.
In this configuration, the conductance is identical for both chiralities, highlighting the essential role of the AM electrode.
These results conclusively demonstrate that the MR observed in our system is determined by the combination of the CISS effect in the molecule and the momentum-dependent spin splitting in the AM.
Moreover, dephasing is also a crucial element \cite{Zhang2025b}.
The influence of dephasing on MR is discussed and investigated in detail in Appendix \ref{secC} and Fig.~\ref{figC1} therein.
It shows that a significant MR persists over a wide range of parameters.

\begin{figure}[t]
\centering
\includegraphics[width = 1.0 \linewidth]{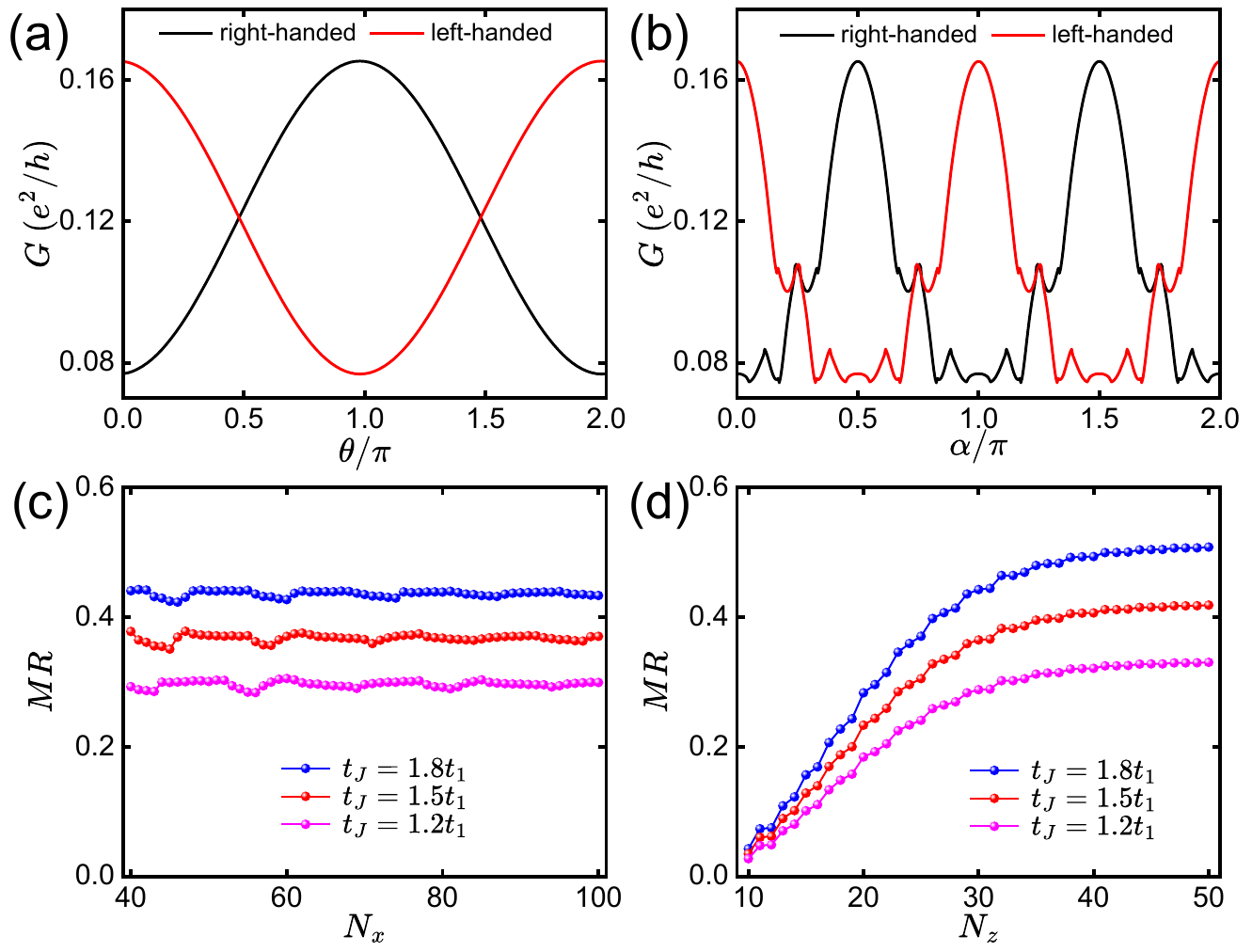}
\caption{\label{fig3}
(a,b) Conductance as a function of (a) the Néel vector angle $\theta$ and (b) the crystal orientation angle $\alpha$, for right-handed (black) and left-handed (red) molecules. In (a), $\alpha=0$ and in (b), $\theta=0$.
(c,d) MR versus the electrode width $N_x$ (c) and molecular length $N_z$ (d) under different AM strength $t_J$.
 In (a-d), $\mu=-1.27t_1$, and $E=t_1$.
 In (c,d), the chirality of the molecule is right-handed.
}
\end{figure}

We now study the effect of the AM parameters, with $\mu=-1.27t_1$ and $E=t_1$.
Figure \ref{fig3}(a) shows the dependence of conductance on the orientation of Néel vector $\theta$ with $\alpha=0$.
The conductances of both right- and left-handed molecules exhibit a period of $2\pi$, because the spin texture on the AM's Fermi surface recovers its original state after a full $2\pi$ rotation.
We begin by examining the conductance when molecule is right-handed.
When $\theta=0$, the conductance is small, because the spin orientation favored to flow in AM mismatches the spin polarization direction of the chiral molecule.
As $\theta$ increases from $0$ to $\pi$, the favored spin orientation of AM gradually matches the molecular spin polarization, leading to an increase in conductance.
With a further increase in $\theta$, the two orientations become mismatched again, causing the conductance to decrease.
For the left-handed molecule, the spin-polarized direction is opposite.
Consequently, at $\theta=0$ the polarization is aligned with AM, giving a larger conductance; at $\theta=\pi$ they are mismatched, resulting in a smaller conductance.
As a result, the conductance of the left-handed molecule is shifted by $\pi$ relative to that of the right-handed molecule, which can be expressed as $G(\theta,-\Delta \varphi)=G(\theta+\pi,\Delta \varphi)$.
These results show that the Néel vector in AM serves as an effective switch for precisely tuning the transport behavior.

Figure \ref{fig3}(b) shows the dependence of conductance on the crystal orientation angle $\alpha$ with $\theta=0$.
The conductances of both right- and left-handed molecules exhibit a period of $\pi$, differing from the $2\pi$ period observed in Fig. \ref{fig3}(a).
This periodicity originates from the symmetry of the AM's Fermi surface, which recovers its original state after a rotation of $\alpha=\pi$ \cite{SunYF2025}.
We first study the conductance of the case of a right-handed molecule.
When $\alpha=0$, the conductance is small because of the mismatch between the favored spin orientation in the AM and the spin polarization of the chiral molecule.
In contrast, at $\alpha=0.5\pi$, the conductance becomes large as these two spin directions are matched.
As $\alpha$ increases from 0 to $0.5\pi$, the conductance tends to increase, and as $\alpha$ increases from $0.5\pi$ to $\pi$, the conductance tends to decrease in general.
The conductance variation with the crystal orientation angle $\alpha$ demonstrates the anisotropic nature of the AM's Fermi surface.
Furthermore, the conductance $G$ is symmetric about $\alpha=0.5\pi$, which can be explained as follows.
The transport of the system is along the $z$ direction, and the transmission probability depends on the $z$ axis projection of the AM Fermi surface.
It is clear that the AM Fermi surface has identical $z$ axis projections in orientation $\alpha$ and $\pi-\alpha$, showing the same transmission probability \cite{SunYF2025}.
As a result, $G(\alpha)=G(\pi-\alpha)$ holds.
The conductance of the left-handed molecule exhibits the same functional dependence on $\alpha$ as the right-handed one, except that its phase is shifted by $0.5\pi$.
This is because in the left-handed case, the spin polarization of the injected current is reversed.
To recover the original spin alignment, the Fermi surface of the AM must be rotated by $\alpha=0.5\pi$, which leads to the relation $G(\alpha,-\Delta \varphi)=G(\alpha+0.5\pi,\Delta \varphi)$.
At $\alpha = 0.25\pi$, $0.75\pi$, $1.25\pi$ and $1.75\pi$, the two curves intersect.
This is because at these values, the projections of the AM Fermi surface onto the $z$-axis are identical, exhibiting no spin splitting along this direction.
Therefore, the conductances become identical for opposite molecular chiralities.
These results demonstrate that the crystal orientation in the AM has a direct impact on the transport properties.

We then investigate the size dependence of MR on the electrode width $N_x$, the molecular length $N_z$, and AM strength $t_J$.
Other parameters are fixed at $\mu=-1.27t_1$, $E=t_1$, and $\alpha=0$.
The chirality of the molecule is right-handed.
Figure \ref{fig3}(c) shows the variation of MR with $N_x$ for several values of AM parameter $t_J$, while fixing $N_z=30$.
The results manifest that the MR of the system is nearly independent of $N_x$, indicating that the effect originates from the CISS of a single molecule and is not influenced by the lateral dimensions of the electrode.
The MR curve also shows small piecewise continuous variation, which arises from the transverse modes at a finite width and diminishes for larger electrode widths.
Figure \ref{fig3}(d) shows the MR as a function of $N_z$ by fixing $N_x=41$.
Here, the MR initially increases with $N_z$ before saturating at longer lengths.
This characteristic dependence is well-aligned with previous theoretical and experimental studies on the CISS effect \cite{Guo2012,Guo2014,Mishra2020}.
It is well-established that spin polarization in CISS builds up with increasing molecular length, which naturally leads to the enhancement of MR \cite{Zhang2025a}.
In both Figs. \ref{fig3}(c,d), the MR increases consistently with $t_J$.
This is because as $t_J$ increases, the spin splitting in the AM becomes more pronounced.
This allows the spin-polarized current to transmit more easily at a certain orientation of the Néel vector, while hindering its transmission for a reversed Néel vector.
The increased difference in $G$ between the two configurations therefore directly leads to a larger MR.

\begin{figure}[t]
\centering
\includegraphics[width = 1.0 \linewidth]{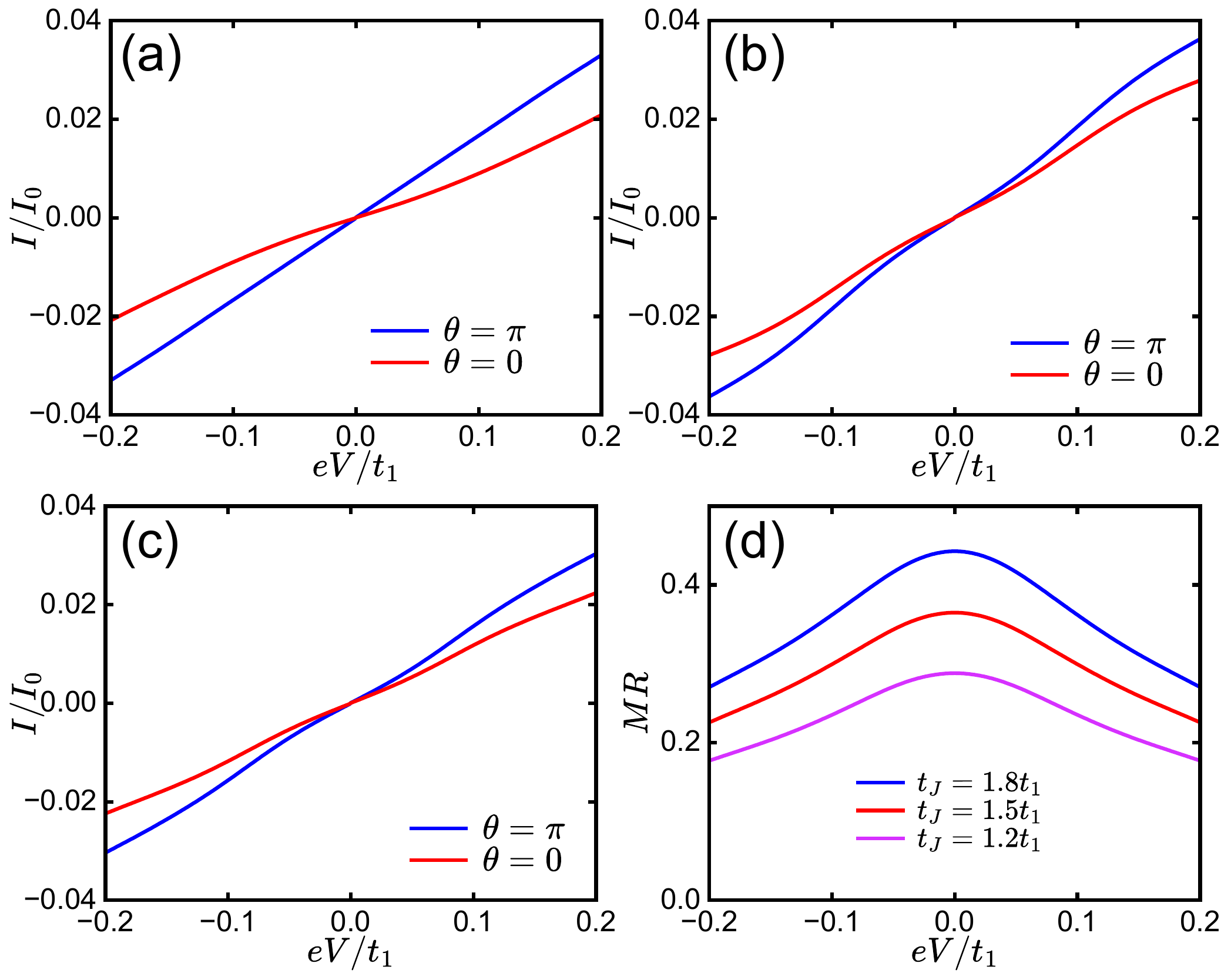}
\caption{\label{fig4}
I-V results for finite bias. 
(a-c) Currents of the device $I$ versus the voltage $V$ at 
$\mu=-1.27t_1$ (a), $\mu=-1.4t_1$ (b), and $\mu=-1.2t_1$ (c),
for $\theta=\pi$ (blue line) and $\theta=0$ (red line), with $t_J=1.5t_1$.
These results show that the MR effect in the device persists over a wide range of chemical potentials.
(d) MR versus the voltage $V$ for different $t_J$ at $\mu=-1.27t_1$.
The chirality of the molecule is right-handed.
The parameters are $\alpha=0$ and $E=t_1$.
$I_0=et_1/h$ is the current unit.
}
\end{figure}

\section{\label{SEC4} MR in finite bias}

Finally, to model realistic experimental conditions with finite bias, we compute the current through the system at different bias voltages.
The current flowing across the device can be calculated by \cite{Zhang2025b} $I=\frac{1}{e}\int_{E-eV/2}^{E+eV/2} G(\varepsilon) \mathrm{d}\varepsilon $ with $V$ the voltage.
Then in finite bias, $MR(V)=[I(\theta=\pi)-I(\theta=0)]/[I(\theta=\pi)+I(\theta=0)]$.
Figures \ref{fig4}(a-c) show the currents $I(\theta=\pi)$ and $I(\theta=0)$ versus $V$ for various chemical potentials.
As the magnitude of voltage $V$ increases, the I-V curves exhibit nonlinear and antisymmetric behavior.
A clear difference between the two I-V curves is observed over a wide voltage range.
Their shape closely resembles the characteristic curves measured using FM electrodes to detect the CISS effect \cite{Bloom2024}, even though the entire present system lacks any macroscopic magnetism.
Figure \ref{fig4}(d) shows the MR derived from Fig. \ref{fig4}(a). The MR exhibits a symmetric lineshape and enhances with increasing $t_J$.
These results demonstrate that a significant MR effect can be generated in the system under realistic finite bias.

\section{\label{SEC5} Conclusion}

In conclusion, we propose the concept of an unconventional spin valve that integrates a chiral molecule with AM, which overcomes the shortcomings of conventional spin valves based on FMs.
This new type of spin valve features high molecular spin polarization, the absence of macroscopic magnetization, and fast response characteristics.
Using a standard numerical approach, we demonstrate that a considerable MR can indeed be achieved in the NM/chiral molecule/AM junction, and its conductance can be precisely and conveniently tuned by controlling the orientation of the AM Néel vector.
Our results promote the interconnection of both the CISS and AM fields, and open a new avenue for the development of spintronic devices.

\section*{ACKNOWLEDGMENTS}
This work was financially supported by the National Key R and D Program of China (Grant No. 2024YFA1409002),
the National Natural Science Foundation of China (Grant No. 12374034, No. 12274466 and No. 125B2065),
Quantum Science and Technology-National Science and Technology Major Project (2021ZD0302403),
and Hunan Provincial Science Fund for Distinguished Young Scholars (Grant No. 2023JJ10058).
We also acknowledge the High-performance Computing Platform of Peking University for providing computational resources.

\newpage

\appendix

\begin{widetext}

\section{\label{secA} \uppercase{Details of Model Hamiltonian}}
\setcounter{equation}{0}
\setcounter{figure}{0}
\renewcommand{\theequation}{A\arabic{equation}}
\renewcommand{\thefigure}{A\arabic{figure}}

The discrete Hamiltonian of the AM electrode can be written as \cite{Cheng2024,SunYF2025}:

\begin{equation}
    \begin{aligned}
    H_{AM}=& 4 t_0 \sum_{\substack{1 \leqslant n_x \leqslant N_x \\ n_z \geqslant N_z + 1}} a_{n_x,n_z}^{\dagger}a_{n_x,n_z}
    -\sum_{\substack{1 \leqslant n_x \leqslant N_x - 1 \\ n_z \geqslant N_z + 1}} (a_{n_x,n_z}^{\dagger}[t_0 - t_J \cos (2\alpha) \bm{\sigma} \cdot \bm{s}]a_{n_x + 1,n_z} + \text{H.c.}) \\
    & -\sum_{\substack{1 \leqslant n_x \leqslant N_x \\ n_z \geqslant N_z + 1}} (a_{n_x,n_z}^{\dagger}[t_0 + t_J \cos (2\alpha) \bm{\sigma} \cdot \bm{s}]a_{n_x,n_z + 1} + \text{H.c.}) \\
    &-\sum_{\substack{1 \leqslant n_x \leqslant N_x - 1 \\ n_z \geqslant N_z + 1}} (a_{n_x,n_z}^{\dagger}[t_J \sin (2\alpha) \bm{\sigma} \cdot \bm{s} / 2]a_{n_x + 1,n_z + 1} + \text{H.c.}) \\
    & +\sum_{\substack{2 \leqslant n_x \leqslant N_x \\ n_z \geqslant N_z + 1}} (a_{n_x,n_z}^{\dagger}[t_J \sin (2\alpha) \bm{\sigma} \cdot \bm{s} / 2]a_{n_x - 1,n_z + 1} + \text{H.c.}),
    \end{aligned}
\label{HAM}
\end{equation}
\end{widetext}
where $a_{n_x,n_z}^{\dagger}=(a_{n_x,n_z,\uparrow}^{\dagger},a_{n_x,n_z,\downarrow}^{\dagger})$ with $n_z \geqslant N_z + 1$ is the creation operator on lattice site $(n_x, n_z)$ in the AM electrode with width $N_x$.

The discrete Hamiltonian of the NM electrode is:
\begin{equation}
    \begin{aligned}
    H_{NM}&=4 t_0 \sum_{\substack{1 \leqslant n_x \leqslant N_x \\ n_z \leqslant 0}} a_{n_x,n_z}^{\dagger}a_{n_x,n_z}\\
    &-t_0 \sum_{\substack{1 \leqslant n_x \leqslant N_x - 1 \\ n_z \leqslant 0}} (a_{n_x,n_z}^{\dagger}a_{n_x + 1,n_z}+\text{H.c.}) \\
    &- t_0 \sum_{\substack{1 \leqslant n_x \leqslant N_x \\ n_z \leqslant -1}} (a_{n_x,n_z}^{\dagger}a_{n_x,n_z + 1} + \text{H.c.}),
    \end{aligned}
\label{HNM}
\end{equation}
where $a_{n_x,n_z}^{\dagger}$ with $n_z \leqslant 0$ is the creation operator on lattice site $(n_x, n_z)$ in the NM electrode with width $N_x$.

The coupling Hamiltonian between electrodes and chiral molecule is written as: $H_{T}=t_c c_{N_z}^{\dagger}a_{N_m,N_z + 1}+t_c c_{1}^{\dagger}a_{N_m,0}+\text{H.c.}$, where $N_m = \left[(N_x+1)/2\right]$, and $\left[\cdot\right]$ is the integer ceiling function.
This indicates that the molecule is connected at the middle of the electrodes with coupling energy $t_c$.

The molecular model presented in Eq.~\ref{Hcmol} does not include interactions; this is entirely justified.
In chiral protein molecules, each amino acid hosts many energy levels.
When the interactions among these levels are taken into account, the molecular Hamiltonian can be written as $\sum_{n_z}(\sum_{l} \varepsilon_{n_z l}c_{n_z l}^{\dagger}c_{n_z l}+\sum_{l,l'} U_{l l'} c_{n_z l}^{\dagger}c_{n_z l}c_{n_z l'}^{\dagger}c_{n_z l'})$, where $\varepsilon_{n_z l}$ denotes the on-site energy of the amino acid, $U_{l l'}$ is the interaction strength, $n_z$ labels the amino acid position, and $l$ ($l'$) labels the orbital on each amino acid.
When an external voltage is applied, electronic transport predominantly occurs near the Fermi level.
For energy levels above the Fermi level, there is no electron occupation; for those below, the levels are fully occupied.
Consequently, apart from the level at the Fermi level, the electron occupation numbers $\left<c_{n_z l'}^{\dagger}c_{n_z l'} \right>$ on all other levels exhibit very small fluctuations, and only one energy level nearest the Fermi level in each amino acid is active.
In this case, the mean-field theory works well, and the system can be described by $\sum_{n_z}\sum_{l} (\varepsilon_{n_z l}c_{n_z l}^{\dagger}c_{n_z l}+\sum_{l'} U_{l l'} c_{n_z l}^{\dagger}c_{n_z l}\left<c_{n_z l'}^{\dagger}c_{n_z l'}\right>)$, which is equivalent to renormalizing the on-site energy of the Fermi-level orbital as $\varepsilon_{n_z l} \to \varepsilon_{n_z l}+\sum_{l'} U_{l l'} \left<c_{n_z l'}^{\dagger}c_{n_z l'}\right>$ \cite{Stafford1996,Yeyati1997,Sun2002}, without altering any physics.
Furthermore, modeling molecular transport using a noninteracting model is a very common and well-established approach that has proven highly effective \cite{Chakraborty2007}.
Our model is therefore well justified.

\section{\label{secB} \uppercase{Transport formulation and spin polarization}}
\setcounter{equation}{0}
\setcounter{figure}{0}
\renewcommand{\theequation}{B\arabic{equation}}
\renewcommand{\thefigure}{B\arabic{figure}}

\begin{figure}[b]
\centering
\includegraphics[width = 1.0 \linewidth]{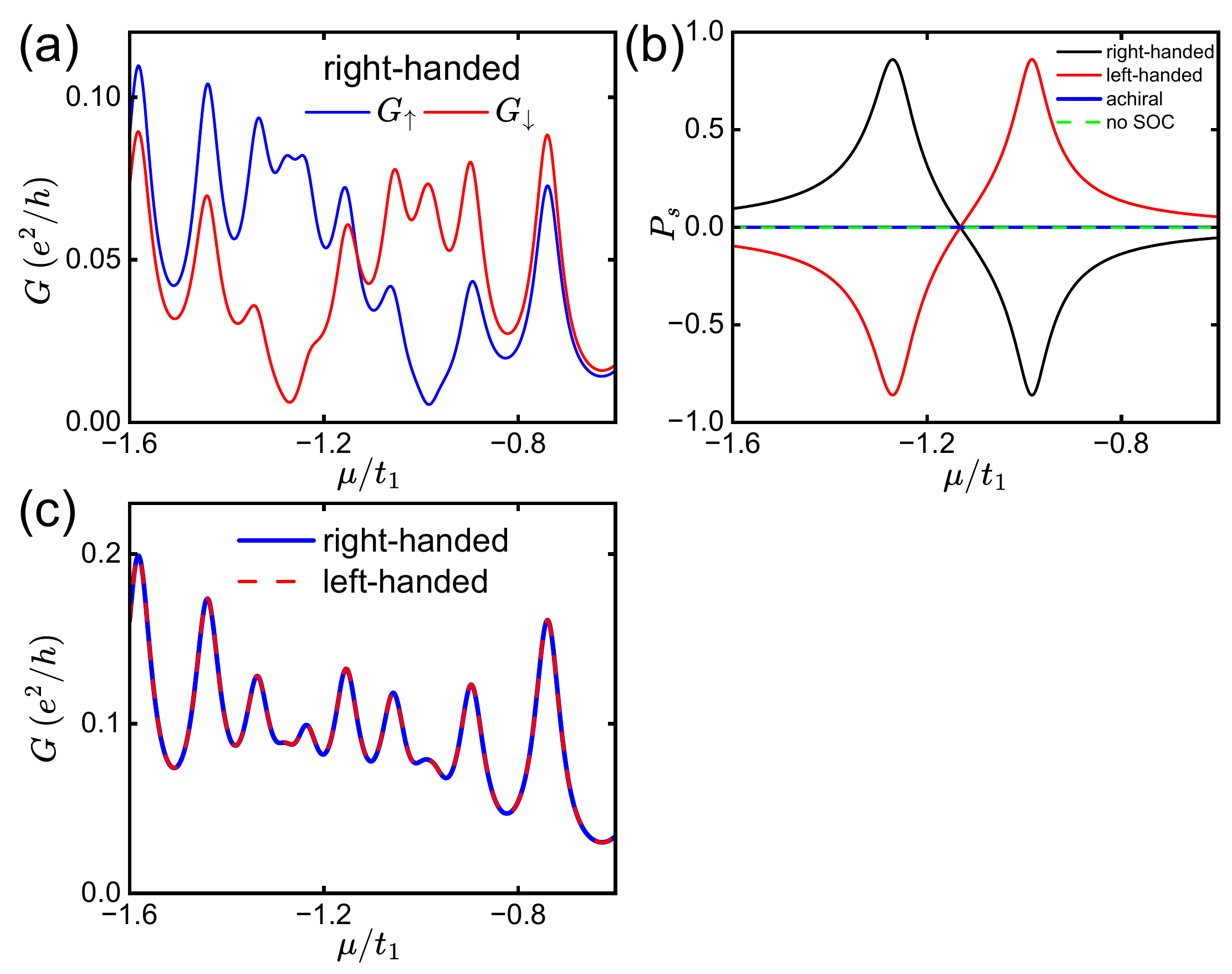}
\caption{\label{figB1}
The conductance of NM/chiral molecule/NM device. 
(a) Conductances $G_{\uparrow}$ and $G_{\downarrow}$ versus the chemical potential $\mu$ in right-handed molecule. 
(b) Spin polarization $P_s$ versus the chemical potential $\mu$. 
(c) Total conductance of the NM/chiral molecule/NM junction. Other parameters are the same as that in Fig.~\ref{fig2}(a).}
\end{figure}

\begin{figure}[t]
\centering
\includegraphics[width = 1.0 \linewidth]{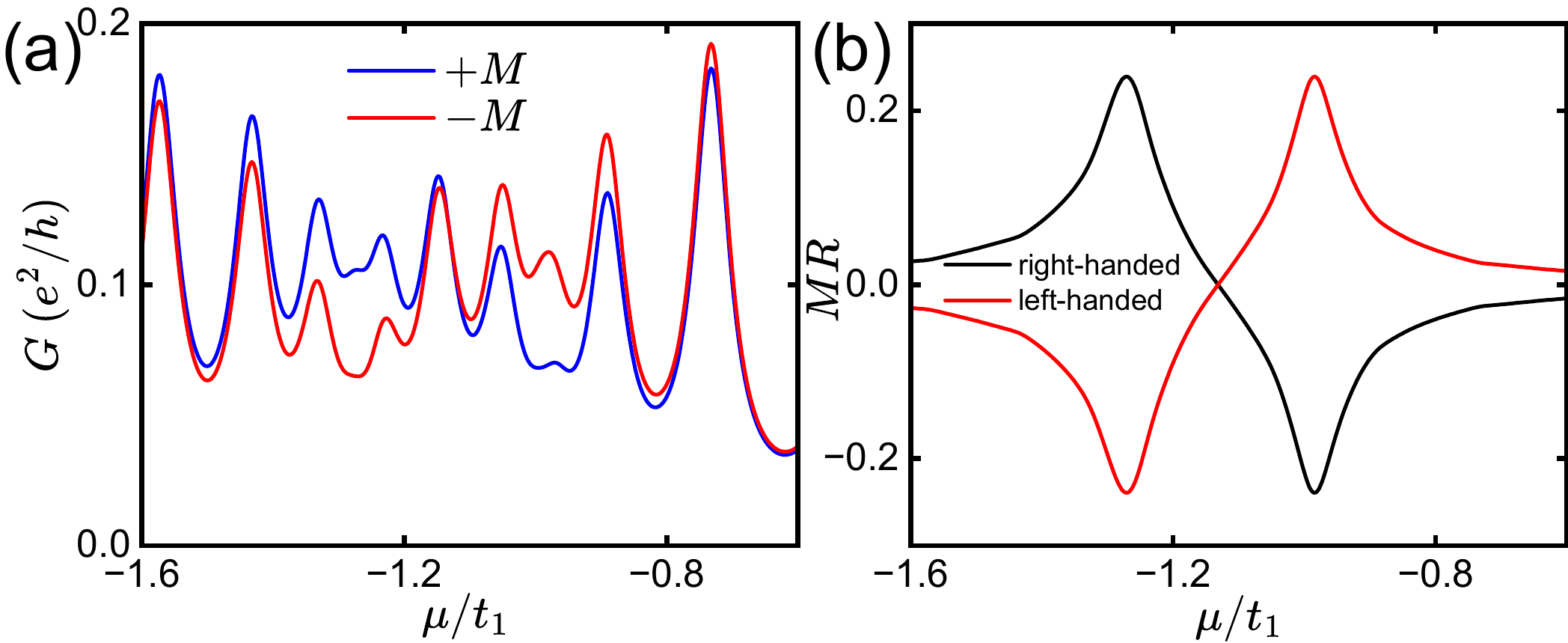}
\caption{\label{figB2}
(a) Conductance $G$ versus chemical potential $\mu$ in NM/right-handed molecule/FM device. 
Blue and red lines are the conductances when the FM electrode is positively magnetized [$G(+M)$] and negatively magnetized [$G(-M)$], respectively. 
(b) MR of the device. The FM polarization is fixed at $P_{FM}=30\%$.}
\end{figure}

In this appendix, we present the formulae for calculating the conductances.
For the case of the NM/chiral molecule/AM junction, according to the Landauer-Büttiker formula \cite{Datta1995}, the conductance is $G=(e^2/h) \sum_{m} T_{AM, m} V_m/V_b$.
Here, $T_{AM, m}=\text{Tr}[\mathbf{\Gamma}_{AM}\mathbf{G}^r\mathbf{\Gamma}_{m}\mathbf{G}^a]$ is the transmission coefficient from the $m$-th electrode to the AM electrode.
The Green's function $\mathbf{G}^r=[\mathbf{G}^a]^{\dagger}=[E \mathbf{I}-\mathbf{H}_{cmol}-\sum_{m}\mathbf{\Sigma}_{m}^r]^{-1}$, and the linewidth function $\mathbf{\Gamma}_{m}=\text{i} [\mathbf{\Sigma}_{m}^r-\mathbf{\Sigma}_{m}^a]$, where the bold letters represent the matrix under the lattice site representation.
$E$ is the Fermi energy, $\mathbf{I}$ is the identity matrix, and $\mathbf{\Sigma}_{m}^r$ is the retarded self-energy introduced by the coupling of the $m$-th electrode ($m= \text{NM}, \text{AM}, 1, 2, \cdots N_z$, representing the NM, AM, 1st, 2nd, $\cdots$ $N_z$-th electrodes).
For the NM and AM electrodes, the self-energy can be calculated numerically \cite{Lee1981}.
For the virtual electrodes, $\mathbf{\Sigma}_{m}^r=-\text{i}\Gamma_d/2=-\text{i}\pi \rho_d t_d^2$ is energy independent, with $\rho_d$ being the density of states in each virtual electrode and $\Gamma_d=0.06t_1$ the dephasing parameter.
$V_m$ is the voltage of the $m$-th virtual electrode, and $V_b$ is a small bias between the NM and AM electrodes with $V_{NM}=V_b, \ V_{AM}=0$. 
Considering the occurrence of the dephasing of electron transport in molecules and the influence of these real electrodes, the voltage on the first and the last $N_d$ virtual electrodes are set to be the same as that of the adjacent real electrodes, and here we set $N_d=\left[N_z/3\right]$. 
The voltage linearly decreases in the middle area because of the uniform distribution of molecular sites.

For the case of the NM/chiral molecule/NM junction, the conductance for spin-up ($G_{\uparrow}$) and spin-down ($G_{\downarrow}$) electrons can be calculated by $G_s=(e^2/h)\sum_{m,s'}T_{Rs,ms'} V_m/V_b$, where $T_{Rs,ms'}=\text{Tr}[\mathbf{\Gamma}_{Rs}\mathbf{G}^r\mathbf{\Gamma}_{ms'}\mathbf{G}^a]$ is the transmission coefficient from the $m$-th electrode with spin $s'=\uparrow, \downarrow$ to the right electrode with spin $s=\uparrow, \downarrow$.
Here, the spin-up conductance $G_{\uparrow}$ and spin-down conductance $G_{\downarrow}$ are proportional to the number of spin-up and spin-down electrons emitted from the right NM electrode.
$V_b$ is the bias voltage, and $V_m$ is the voltage of the $m$-th electrode.
The Green's function $\mathbf{G}^r=[\mathbf{G}^a]^{\dagger}=[E \mathbf{I}-\mathbf{H}_{cmol}-\sum_{ms}\mathbf{\Sigma}_{ms}^r]^{-1}$, and $\mathbf{\Gamma}_{ms}=\text{i} [\mathbf{\Sigma}_{ms}^r-\mathbf{\Sigma}_{ms}^a]$. 
Here $\mathbf{\Sigma}_{ms}^r$ is the retarded self-energy due to the coupling to the $m$-th electrode, and $m=L, R, 1, 2, \cdots, N_z$ represent the left, right, virtual 1st, 2nd …, $N_z$-th electrodes.
Here, all settings for the electrodes are identical to those in the NM/chiral molecule/AM junction, except that the right electrode is replaced with an NM electrode.
Finally, the spin polarization is defined as $P_s=(G_{\uparrow}-G_{\downarrow})/(G_{\uparrow}+G_{\downarrow})$.

For the case of the NM/chiral molecule/FM junction, the conductance is $G=(e^2/h)\sum_{m}T_{FM,m} V_m/V_b$.
Here, $T_{FM,m}=\text{Tr}[\mathbf{\Gamma}_{FM}\mathbf{G}^r\mathbf{\Gamma}_{m}\mathbf{G}^a]$ is the transmission coefficient from the $m$-th electrode to the FM electrode, with $\mathbf{\Gamma}_{FM,\uparrow}=(1+P_{FM})\mathbf{\Gamma}_{NM,\uparrow}$, $\mathbf{\Gamma}_{FM,\downarrow}=(1-P_{FM})\mathbf{\Gamma}_{NM,\downarrow}$ and $\mathbf{\Sigma}_{FM}^r=-\text{i}\mathbf{\Gamma}_{FM}/2$.
Here, $P_{FM}$ is the spin polarization of the FM, and we take $P_{FM}=30\%$, corresponding to a typical value for a common FM \cite{Zhang2025b}.
All other configurations and computational details remain identical to those in the AM case. 

The conductances for different spins of the NM/chiral molecule/NM junction are shown in Fig.~\ref{figB1}(a).
It shows that $G_{\uparrow}$ differs significantly from $G_{\downarrow}$ over a wide chemical potential range, indicating that spin-unpolarized incident electrons from the left NM electrode become highly spin-polarized after transmitting through the chiral molecule toward the right NM electrode.
Figure \ref{figB1}(b) displays the spin polarization $P_s$ for right-handed, left-handed, achiral, and no SOC molecular junctions.
The chiral molecules exhibit high spin polarization with opposite signs for opposite chiralities, whereas both the achiral and no SOC molecules yield zero spin polarization. These features are clear signatures of the CISS effect.

Figure \ref{figB2} shows the conductance $G$ of the NM/chiral molecule/FM junction as a function of chemical potential $\mu$ for the positive and negative magnetization directions.
The conductance for positive magnetization $G(+M)$ and that for negative magnetization $G(-M)$ exhibit a clear difference.
The magnitude of the conductance variation is slightly reduced compared to that in the AM electrode case presented in Fig.~\ref{fig2} in the main text.

Figure \ref{figB2}(b) shows the MR of the NM/chiral molecule/FM system, defined as $MR=[G(+M)-G(-M)]/[G(+M)+G(-M)]$, as a function of chemical potential $\mu$ \cite{Zhang2025b}.
The MR curve for the FM electrode case shows a similar shape to that for the AM electrode case, but with a reduced magnitude (the maximum MR is 23.9\% for the FM electrode case, compared to 36.5\% for the AM electrode case).
This is because, in this device, the transport direction is along the $z$-axis, along which the AM exhibits strong spin splitting, and consequently, a strong spin polarization is present in the transport direction \cite{ZhangX2025}.
This spin polarization can be stronger than that in FM, so the MR of the AM could be larger. 
Experimentally, the actual measured MR depends on both the specific material and the transport direction.

\section{\label{secC} \uppercase{The effect of dephasing}}
\setcounter{equation}{0}
\setcounter{figure}{0}
\renewcommand{\theequation}{C\arabic{equation}}
\renewcommand{\thefigure}{C\arabic{figure}}

\begin{figure}[t]
\centering
\includegraphics[width = 1.0 \linewidth]{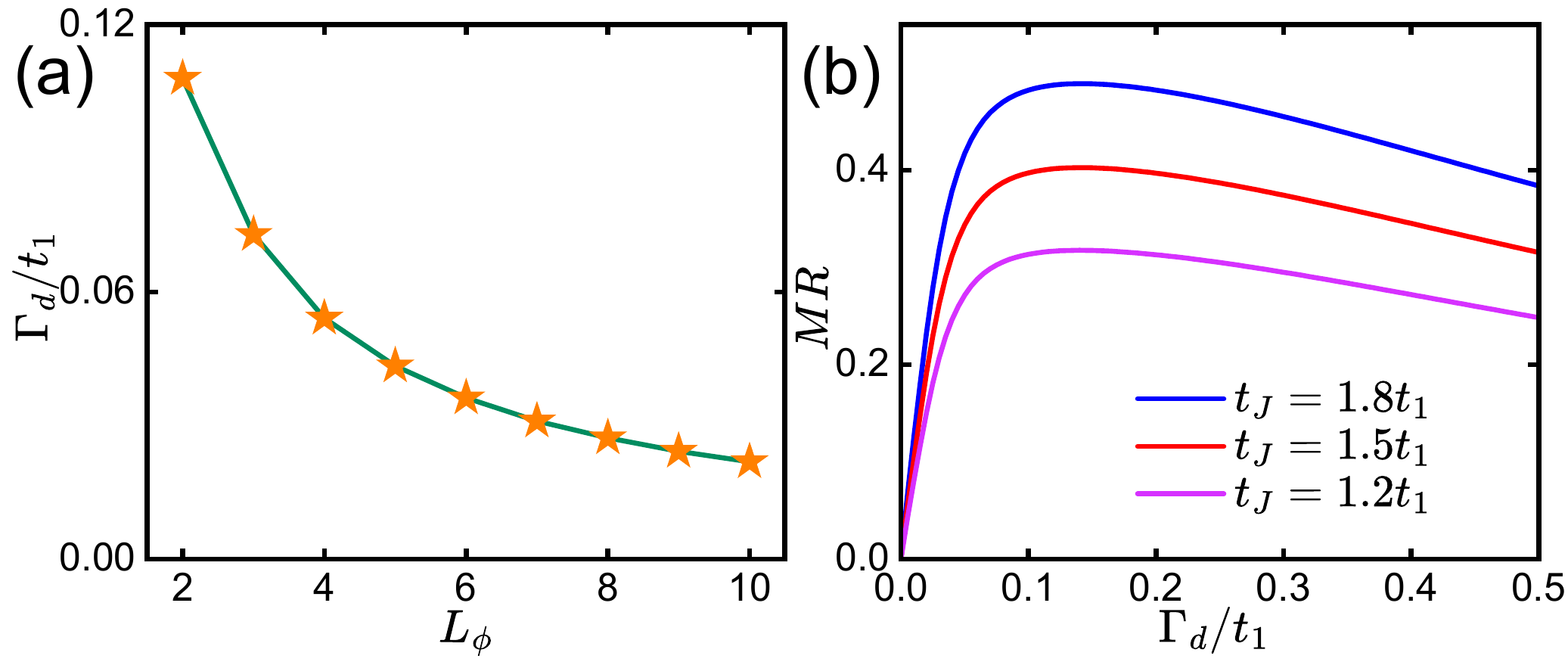}
\caption{\label{figC1}
The effect of dephasing. 
(a) The conversion relationship between $L_{\phi}$ (in the unit of sites) and $\Gamma_d$. 
(b) MR vs. the dephasing strength $\Gamma_d$. 
Other parameters are the same as that in Fig.~\ref{fig2}(a).}
\end{figure}

Dephasing plays a crucial role in the CISS effect. 
In this appendix, we discuss and investigate the influence of dephasing on the transport results.
First, we discuss the role of dephasing on the results. 
In chiral molecules, SOC and the helical structure cause a difference in the propagation velocities between spin-up and spin-down electrons, while dephasing leads to the loss of both phase and spin memory \cite{Zhang2025a}.
Consider an equal number of spin-up and spin-down electrons entering the chiral molecule, with the spin-up electrons traveling faster.
In the absence of dephasing ($\Gamma_d=0$), although electrons with different spins propagate at different velocities, the numbers of spin-up and spin-down electrons that eventually traverse the molecule remain equal.
Hence, no net spin polarization emerges, and consequently no MR is produced.
When dephasing is present, the faster spin-up electrons experience less dephasing as they propagate quickly through the molecule, largely preserving their spin orientation.
In contrast, the slower spin-down electrons undergo more dephasing, which can cause them to lose spin memory.
Once flipped to spin-up, these electrons have faster velocities and leave the molecule rapidly; if they remain spin-down, they continue to move slowly and are subject to further dephasing \cite{Zhang2025a}.
This process ultimately leads to an increase in the number of spin-up electrons, generating a net spin polarization and a finite MR.
However, when dephasing becomes too strong, both spin species are strongly affected, and the resulting spin polarization as well as the MR are suppressed.

We then discuss the physical meaning and realism of the dephasing parameter $\Gamma_d$.
In molecular systems, dephasing arises from phase-breaking processes such as inelastic scattering by electrons, impurities, adsorbed counterions, or phonons, all of which cause electrons to lose their coherence.
The strength $\Gamma_d$ quantifies the rate at which electrons lose coherence. 
A larger $\Gamma_d$ corresponds to faster phase loss and thus a shorter coherence length $L_{\phi}$.
To illustrate this, we calculate the relationship of $L_{\phi}$ and $\Gamma_d$, shown in Fig.~\ref{figC1}(a).
The calculation method is as follows \cite{Xing2008}: A molecule with length $L_z$ is contacted with two real left and right electrodes (labeled L and R), and the structure of these electrodes is identical to that of the central molecular region, ensuring no reflection at the contact interfaces.
The transmission coefficients from electrode-$q$ to electrode-$p$ $T_{pq}$ can be calculated with the same method we described in Appendix~\ref{secB}.
Here, $p,q=L,R,1,2,\cdots,L_z$.
Thus, all transmission coefficients are functions of $L_z$ and $\Gamma_d$ [$T_{pq}=T_{pq} (L_z,\Gamma_d)$].
$T_{RL}$ represents the probability that electrons go through the molecule without dephasing.
The summation of transmission coefficients from the real left electrode to all the virtual electrodes $\sum_{v=1,2,\cdots}T_{vL}$ represents the probability that electrons lose their phase information.
If we fix $L_z$ and increase $\Gamma_d$, $T_{RL}$ will decrease and $\sum_{v}T_{vL}$ will increase.
When $T_{RL} (L_z,\Gamma_d)=\sum_v T_{vL} (L_z,\Gamma_d)$, which means the probability of an electron from the left electrode losing and keeping phase information are both 50\%, $L_z=L_{\phi}$.
Therefore, the relationship between $L_{\phi}$ and $\Gamma_d$ is obtained by numerically solving $T_{RL} (L_{\phi},\Gamma_d)=\sum_v T_{vL} (L_{\phi},\Gamma_d)$.

It can be observed in Fig.~\ref{figC1}(a) that a smaller $\Gamma_d$ indeed yields a longer $L_{\phi}$.
The value used in our work, $\Gamma_d=0.06t_1$, corresponds to a coherence length of $L_{\phi} \approx 3 \sim 4$ (equivalent to 1.4-1.8 nm).
Experimentally, it has been shown that the phase coherence length in helical molecules is well below 4 nm \cite{Morita2003}.
Therefore, our choice of $\Gamma_d$ falls within a physically reasonable and experimentally supported range.

Finally, we explore the dependence of MR on $\Gamma_d$.
As shown in Fig.~\ref{figC1}(b), the MR initially increases with $\Gamma_d$, and then slowly declines upon further increase $\Gamma_d$, which is fully consistent with the analysis above.
Moreover, a significant MR persists over a wide range of $\Gamma_d$ values, demonstrating the broad applicability of our model.

\bibliography{mainbibs}

@PREAMBLE{
 "\providecommand{\noopsort}[1]{}" 
 # "\providecommand{\singleletter}[1]{#1}%" 
}

@ARTICLE{Hsieh2009,
   author       = "D. Hsieh and Y. Xia and L. Wray and D. Qian and A. Pal and J. H. Dil and J. Osterwalder and F. Meier and G. Bihlmayer and C. L. Kane and others",
   year         = "2009",
   journal      = "Science",
   volume       = "323",
   pages        = "919",
   title        = "Observation of unconventional quantum spin textures in topological insulators",
   doi          = "10.1126/science.1167733",
}

@ARTICLE{Bornscheuer2012,
   author       = "U. T. Bornscheuer and G. W. Huisman and R. J. Kazlauskas and S. Lutz and J. C. Moore and K. Robins", 
   year         = "2012",
   journal      = "Nature",
   volume       = "485",
   pages        = "185",
   title        = "Engineering the third wave of biocatalysis",
   doi = "https://doi.org/10.1038/nature11117",
}

@ARTICLE{Ray1999,
   author       = "K. Ray and S. P. Ananthavel and D. H. Waldeck and R. Naaman", 
   year         = "1999",
   journal      = "Science",
   volume       = "283",
   pages        = "814",
   title        = "Asymmetric scattering of polarized electrons by organized organic films of chiral molecules",
   doi = "10.1126/science.283.5403.814", 
}

@ARTICLE{Xu2023,
   author       = "Y. Xu and W. Mi", 
   year         = "2023",
   journal      = "Mater. Horiz.",
   volume       = "10",
   pages        = "1924",
   title        = "Chiral-induced spin selectivity in biomolecules, hybrid organic-inorganic perovskites and inorganic materials: a comprehensive review on recent progress",
   doi          = "10.1039/D3MH00024A"
}

@ARTICLE{Bloom2024,
   author       = "B. P. Bloom and Y. Paltiel and R. Naaman and D. H. Waldeck", 
   year         = "2024",
   journal      = "Chem. Rev.",
   volume       = "124",
   pages        = "1950",
   title        = "Chiral induced spin selectivity",
   doi = "https://doi.org/10.1021/acs.chemrev.3c00661",
}

@ARTICLE{Foo2025,
   author       = "Y. X. Foo and A. Kermiche and F. T. Chowdhury and C. D. Aiello and L. D. Smith", 
   year         = "2025",
   journal      = "Chem. Phys. Rev.",
   volume       = "6",
   pages        = "031306",
   title        = "Mind the gap: From resolving theoretical foundations of chiral(ity)-induced spin selectivity to pioneering implementations in quantum sensing",
   doi = "https://doi.org/10.1063/5.0244306",
}

@ARTICLE{LLi2025,
   author       = "L. Li and W. Shi and A. Mahajan and J. Zhang and M. Gómez-Gómez and J. Labella and S. Louie and T. Torres and S. Barlow and S. R. Marder and others", 
   year         = "2025",
   journal      = "J. Am. Chem. Soc.",
   volume       = "147",
   pages        = "25043",
   title        = "Too fast for spin flipping: absence of chirality-induced spin selectivity in coherent electron transport through single-molecule junctions",
   doi = "https://doi.org/10.1021/jacs.5c08517",
}

@ARTICLE{Gohler2011,
   author       = "B. Göhler and V. Hamelbeck and T. Z. Markus and M. Kettner and G. F. Hanne and Z. Vager and R. Naaman and H. Zacharias", 
   year         = "2011",
   journal      = "Science",
   volume       = "331",
   pages        = "894",
   title        = "Spin selectivity in electron transmission through self-assembled monolayers of double-stranded {DNA}",
   doi = "10.1126/science.1199339",
}

@ARTICLE{Sun2024,
   author       = "R. Sun and K. S. Park and A. H. Comstock and A. McConnell and Y.-C. Chen and P. Zhang and D. Beratan and W. You and A. Hoffmann and Z.-G. Yu and others", 
   year         = "2024",
   journal      = "Nat. Mater.",
   volume       = "23",
   pages        = "782",
   title        = "Inverse chirality-induced spin selectivity effect in chiral assemblies of $\pi$-conjugated polymers",
   doi = "https://doi.org/10.1038/s41563-024-01838-8",
}

@ARTICLE{Wang2024,
   author       = "C. Wang and Z.-R. Liang and X.-F. Chen and A.-M. Guo and G. Ji and Q.-F. Sun and Y. Yan", 
   year         = "2024",
   journal      = "Phys. Rev. Lett.",
   volume       = "133",
   pages        = "108001",
   title        = "Transverse spin selectivity in helical nanofibers prepared without any chiral molecule",
   doi = "https://doi.org/10.1103/PhysRevLett.133.108001",
}

@ARTICLE{Moharana2025,
   author       = "A. Moharana and Y. Kapon and F. Kammerbauer and D. Anthofer and S. Yochelis and H. Shema and E. Gross and M. Kläui and Y. Paltiel and A. Wittmann", 
   year         = "2025",
   journal      = "Sci. Adv.",
   volume       = "11",
   pages        = "1",
   title        = "Chiral-induced unidirectional spin-to-charge conversion",
   doi = "10.1126/sciadv.ado4285",
}

@ARTICLE{Mishra2013,
   author       = "D. Mishra and T. Z. Markus and R. Naaman and M. Kettner and B. Göhler and H. Zacharias and N. Friedman and M. Sheves and C. Fontanesi", 
   year         = "2013",
   journal      = "Proc. Natl. Acad. Sci. U.S.A.",
   volume       = "110",
   pages        = "14872",
   title        = "Spin-dependent electron transmission through bacteriorhodopsin embedded in purple membrane",
   doi = "https://doi.org/10.1073/pnas.1311493110",
}

@ARTICLE{Kiran2016,
   author       = "V. Kiran and S. P. Mathew and S. R. Cohen and I. Hernandez Delgado and J. Lacour and R. Naaman", 
   year         = "2016",
   journal      = "Adv. Mater.",
   volume       = "28",
   pages        = "1957",
   title        = "Helicenes-a new class of organic spin filter",
   doi = "https://doi.org/10.1002/adma.201504725Digital",
}

@ARTICLE{Abendroth2019,
   author       = "J. M. Abendroth and K. M. Cheung and D. M. Stemer and M. S. El Hadri and C. Zhao and E. E. Fullerton and P. S. Weiss", 
   year         = "2019",
   journal      = "J. Am. Chem. Soc.",
   volume       = "141",
   pages        = "3863",
   title        = "Spin-dependent ionization of chiral molecular films",
   doi = "https://doi.org/10.1021/jacs.8b08421",
}

@ARTICLE{Levi2016,
   author       = "M. Eckshtain-Levi and E. Capua and S. Refaely-Abramson and S. Sarkar and Y. Gavrilov and S. P. Mathew and Y. Paltiel and Y. Levy and L. Kronik and R. Naaman", 
   year         = "2016",
   journal      = "Nat. Commun.",
   volume       = "7",
   pages        = "10744",
   title        = "Cold denaturation induces inversion of dipole and spin transfer in chiral peptide monolayers",
   doi = "https://doi.org/10.1038/ncomms10744",
}

@ARTICLE{Kumar2017,
   author       = "A. Kumar and E. Capua and M. K. Kesharwani and J. M. L. Martin and E. Sitbon and D. H. Waldeck and R. Naaman", 
   year         = "2017",
   journal      = "Proc. Natl. Acad. Sci. U.S.A.",
   volume       = "114",
   pages        = "2474",
   title        = "Chirality-induced spin polarization places symmetry constraints on biomolecular interactions",
   doi = "https://doi.org/10.1073/pnas.1611467114",
}

@ARTICLE{Eckvahl2023,
   author       = "H. J. Eckvahl and N. A. Tcyrulnikov and A. Chiesa and J. M. Bradley and R. M. Young and S. Carretta and M. D. Krzyaniak and M. R. Wasielewski", 
   year         = "2023",
   journal      = "Science",
   volume       = "382",
   pages        = "197",
   title        = "Direct observation of chirality-induced spinselectivity in electron donor-acceptor molecules",
   doi = "10.1126/science.adj5328",
}

@ARTICLE{Eckvahl2024,
   author       = "H. J. Eckvahl and G. Copley and R. M. Young and M. D. Krzyaniak and M. R. Wasielewski", 
   year         = "2024",
   journal      = "J. Am. Chem. Soc.",
   volume       = "146",
   pages        = "24125",
   title        = "Detecting chirality-induced spin selectivity in randomly oriented radical pairs photogenerated by hole transfer",
   doi = "https://doi.org/10.1021/jacs.4c08706",
}

@ARTICLE{Latawiec2025,
   author       = "E. I. Latawiec and A. Chiesa and Y. Qiu and N. A. Tcyrulnikov and R. M. Young and S. Carretta and M. D. Krzyaniak and M. R. Wasielewski", 
   year         = "2025",
   journal      = "Proc. Natl. Acad. Sci. U.S.A.",
   volume       = "122",
   pages        = "e2515120122",
   title        = "Detecting chirality- induced spin selectivity in chromophore- linked {DNA} hairpins using photogenerated radical pairs",
   doi = "https://doi.org/10.1073/pnas.2515120122",
}

@ARTICLE{Eckvahl2026,
   author       = "H. J. Eckvahl and N. A. Tcyrulnikov and K. T. Kairys and G. C. Mantel and R. M. Young and M. D. Krzyaniak and M. R. Wasielewski", 
   year         = "2026",
   journal      = "J. Am. Chem. Soc.",
   volume       = "148",
   pages        = "15193",
   title        = "Chirality-induced spin selectivity in an achiral electron donor-acceptor molecule doped into a cholesteric liquid crystal",
   doi = "https://doi.org/10.1021/jacs.6c01757",
}

@ARTICLE{TLiu2024,
   author       = "T. Liu and Y. Adhikari and H. Wang and Y. Jiang and Z. Hua and H. Liu and P. Schlottmann and H. Gao and P. S. Weiss and B. Yan and others", 
   year         = "2024",
   journal      = "Adv. Mater.",
   volume       = "36",
   pages        = "2406347",
   title        = "Chirality-induced magnet-free spin generation in a semiconductor",
   doi = "https://doi.org/10.1002/adma.202406347",
}

@ARTICLE{LLiu2026,
   author       = "L. Liu and P.-Y. Liu and T.-Y. Zhang and Q.-F. Sun", 
   year         = "2026",
   journal      = "Phys. Rev. B",
   volume       = "113",
   pages        = "L121407",
   title        = "Engineering chiral-induced spin selectivity in an artificial topological quantum well",
   doi = "https://doi.org/10.1103/87b7-rq7h",
}

@ARTICLE{Jia2021,
   author       = "L. Jia and C. Wang and Y. Zhang and L. Yang and Y. Yan", 
   year         = "2021",
   journal      = "ACS Nano",
   volume       = "14",
   pages        = "6607",
   title        = "Efficient Spin Selectivity in Self-Assembled Superhelical Conducting Polymer Microfibers",
   doi = "https://doi.org/10.1021/acsnano.9b07681",
}

@ARTICLE{Guo2012,
   author       = "A.-M. Guo and Q.-F. Sun", 
   year         = "2012",
   journal      = "Phys. Rev. Lett.",
   volume       = "108",
   pages        = "218102",
   title        = "Spin-selective transport of electrons in {DNA} double helix",
   doi = "https://doi.org/10.1103/PhysRevLett.108.218102",
}

@ARTICLE{Guo2014,
   author       = "A.-M. Guo and Q.-F. Sun", 
   year         = "2014",
   journal      = "Proc. Natl. Acad. Sci. U.S.A.",
   volume       = "111",
   pages        = "11658",
   title        = "Spin-dependent electron transport in protein-like single-helical molecules",
   doi = "https://doi.org/10.1073/pnas.1407716111",
}

@ARTICLE{Buttiker1986,
   author       = "M. Büttiker", 
   year         = "1986",
   journal      = "Phys. Rev. Lett.",
   volume       = "57",
   pages        = "1761",
   title        = "Four-terminal phase-coherent conductance",
   doi = "https://doi.org/10.1103/PhysRevLett.57.1761",
}

@ARTICLE{Zhang2025a,
   author       = "T.-Y. Zhang and Y. Mao and A.-M. Guo and Q.-F. Sun", 
   year         = "2025",
   journal      = "Phys. Rev. B",
   volume       = "111",
   pages        = "205417",
   title        = "Dynamical theory of chiral-induced spin selectivity in electron donor--chiral molecule--acceptor systems",
   doi = "https://doi.org/10.1103/PhysRevB.111.205417",
}

@ARTICLE{Liu2025a,
   author       = "P.-Y. Liu and T.-Y. Zhang and Q.-F. Sun", 
   year         = "2025",
   journal      = "J. Phys. Chem. Lett.",
   volume       = "16",
   pages        = "6500",
   title        = "Dynamical simulation of chiral-induced spin-polarization and magnetization",
   doi = "https://doi.org/10.1021/acs.jpclett.5c01179",
}

@ARTICLE{Liu2025b,
   author       = "P.-Y. Liu and T.-Y Zhang and A.-M Guo and Y. Paltiel and Q.-F. Sun", 
   year         = "2025",
   journal      = "J. Phys. Chem. Lett.",
   volume       = "16",
   pages        = "10426",
   title        = "Spin-to-charge conversion modulated by chiral molecules",
   doi = "https://doi.org/10.1021/acs.jpclett.5c02519",
}

@ARTICLE{Zhang2025b,
   author       = "T.-Y Zhang and Y. Mao and P.-Y. Liu and A.-M Guo and Q.-F. Sun", 
   year         = "2025",
   journal      = "J. Phys. Chem. Lett.",
   volume       = "16",
   pages        = "12596",
   title        = "Anomalous magnetoresistance beyond the {Jullière} model for spin selectivity in chiral molecules",
   doi = "https://doi.org/10.1021/acs.jpclett.5c03104",
}

@article{Zhang2025c,
  title={Spin-to-charge conversion driven by inverse chiral-induced spin selectivity},
  author={Zhang, Tian-Yi and Liu, Peng-Yi and Guo, Ai-Min and Sun, Qing-Feng},
  journal={arXiv preprint arXiv:2509.04022},
  year={2025},
  note  = {Submission date: 2025-09-04, URL: \url{https://arxiv.org/abs/2509.04022} (Accessed: 2026-02-12)}
}

@ARTICLE{Alwan2021,
   author       = "S. Alwan and Y. Dubi", 
   year         = "2021",
   journal      = "J. Am. Chem. Soc.",
   volume       = "143",
   pages        = "14235",
   title        = "Spinterface origin for the chirality-induced spin-selectivity effect",
   doi = "https://doi.org/10.1021/jacs.1c05637",
}

@ARTICLE{Chiesa2024,
   author       = "A. Chiesa and E. Garlatti and M. Mezzadri and L. Celada and R. Sessoli and M. R. Wasielewski and R. Bittl and P. Santini and S. Carretta", 
   year         = "2024",
   journal      = "Nano Lett.",
   volume       = "24",
   pages        = "12133",
   title        = "Many body models for chirality-induced spin selectivity in electron transfer",
   doi = "https://doi.org/10.1021/acs.nanolett.4c02912",
}

@ARTICLE{Fransson2023,
   author       = "J. Fransson", 
   year         = "2023",
   journal      = "Phys. Rev. Res.",
   volume       = "5",
   pages        = "L022039",
   title        = "Chiral phonon induced spin polarization",
   doi = "https://doi.org/10.1103/PhysRevResearch.5.L022039",
}

@ARTICLE{Dianat2020,
   author       = "A. Dianat and R. Gutierrez and H. Alpern and V. Mujica and A. Ziv and S. Yochelis and O. Millo and Y. Paltiel and G. Cuniberti", 
   year         = "2020",
   journal      = "Nano Lett.",
   volume       = "20",
   pages        = "7077",
   title        = "Role of exchange interactions in the magnetic response and intermolecular recognition of chiral molecules",
   doi = "https://doi.org/10.1021/acs.nanolett.0c02216",
}

@ARTICLE{Chen2024,
   author       = "S. Chen and R. Wu and H.-H. Fu", 
   year         = "2024",
   journal      = "Nano Lett.",
   volume       = "24",
   pages        = "6210",
   title        = "Persistent chirality-induced spin-selectivity effect in circular helix molecules",
   doi = "https://doi.org/10.1021/acs.nanolett.4c00383",
}

@ARTICLE{Sierra2020,
   author       = "M. A. Sierra and D. Sánchez and R. Gutierrez and G. Cuniberti and F. Domínguez-Adame and E. Díaz", 
   year         = "2020",
   journal      = "Biomolecules",
   volume       = "10",
   pages        = "49",
   title        = "Spin-polarized electron transmission in {DNA}-like systems",
   doi = "https://doi.org/10.3390/biom10010049",
}

@ARTICLE{Wang2026,
   author       = "Y. Wang and W. B. Du and X. Liu and J. F. Ren and S. J. Xie and C. Timm and C. K. Wang and G. C. Hu", 
   year         = "2026",
   journal      = "Phys. Rev. B",
   volume       = "113",
   pages        = "104416",
   title        = "Mechanical modulation of spin-polarized transport in chiral molecular junctions",
   doi = "https://doi.org/10.1103/l56v-mspq",
}

@ARTICLE{Kundu2024,
   author       = "S. Kundu and C. Simserides", 
   year         = "2024",
   journal      = "Phys. Rev. E",
   volume       = "109",
   pages        = "014401",
   title        = "Charge transport in a double-stranded {DNA}: Effects of helical symmetry and long-range hopping",
   doi = "https://doi.org/10.1103/PhysRevE.109.014401",
}

@ARTICLE{Sun2026,
   author       = "X. Sun and K.-Y. Zhang and S.-Z. Zhou and H.-H. Fu", 
   year         = "2026",
   journal      = "Nat. Commun.",
   volume       = "17",
   pages        = "1231",
   title        = "Robust chirality-induced spin selectivity in topologically chiral molecular knots",
   doi = "https://doi.org/10.1038/s41467-025-67988-8",
}

@ARTICLE{Adhikari2023,
   author       = "Y. Adhikari and T. Liu and H. Wang and Z. Hua and H. Liu and E. Lochner and P. Schlottmann and B. Yan and J. Zhao and P. Xiong", 
   year         = "2023",
   journal      = "Nat. Commun.",
   volume       = "14",
   pages        = "5163",
   title        = "Interplay of structural chirality, electron spin and topological orbital in chiral molecular spin valves",
   doi = "https://doi.org/10.1038/s41467-023-40884-9",
}

@ARTICLE{Bloom2025,
   author       = "B. P. Bloom and M. Lingenfelder and R. Naaman and D. Sun and D. H. Waldeck", 
   year         = "2025",
   journal      = "Nat. Rev. Mat.",
   volume       = "11",
   pages        = "213",
   title        = "Using chiral-induced spin selectivity as a tool to improve materials and processes for energy science",
   doi  = "https://doi.org/10.1038/s41578-025-00864-5"
}

@ARTICLE{Julliere1975,
   author       = "M. Julliere", 
   year         = "1975",
   journal      = "Phys. Lett. A",
   volume       = "54",
   pages        = "225",
   title        = "Tunneling between ferromagnetic films",
   doi = "https://doi.org/10.1016/0375-9601(75)90174-7",
}

@ARTICLE{Moodera1995,
   author       = "J. S. Moodera and L. R. Kinder and T. M. Wong and R. Meservey", 
   year         = "1995",
   journal      = "Phys. Rev. Lett.",
   volume       = "74",
   pages        = "3273",
   title        = "Large magnetoresistance at room temperature in ferromagnetic thin film tunnel junctions",
   doi = "https://doi.org/10.1103/PhysRevLett.74.3273",
}

@ARTICLE{SunYF2025,
   author       = "Y.-F. Sun and Y. Mao and Y.-C. Zhuang and Q.-F. Sun", 
   year         = "2025",
   journal      = "Phys. Rev. B",
   volume       = "112",
   pages        = "094411",
   title        = "Tunneling magnetoresistance effect in altermagnets",
   doi = "https://doi.org/10.1103/t8b5-l859",
}

@ARTICLE{Bai2024,
   author       = "L. Bai and W. Feng and S. Liu and L. Šmejkal and Y. Mokrousov and Y. Yao", 
   year         = "2024",
   journal      = "Adv. Funct. Mater.",
   volume       = "34",
   pages        = "2409327",
   title        = "Altermagnetism: Exploring new frontiers in magnetism and spintronics",
   doi = "https://doi.org/10.1002/adfm.202409327",
}

@ARTICLE{Smejkal2022a,
   author       = "L. Šmejkal and J. Sinova and T. Jungwirth", 
   year         = "2022",
   journal      = "Phys. Rev. X",
   volume       = "12",
   pages        = "040501",
   title        = "Emerging research landscape of altermagnetism",
   doi = "https://doi.org/10.1103/PhysRevX.12.040501",
}

@ARTICLE{Smejkal2022b,
   author       = "L. Šmejkal and J. Sinova and T. Jungwirth", 
   year         = "2022",
   journal      = "Phys. Rev. X",
   volume       = "12",
   pages        = "031042",
   title        = "Beyond conventional ferromagnetism and antiferromagnetism: A phase with nonrelativistic spin and crystal rotation symmetry",
   doi = "https://doi.org/10.1103/PhysRevX.12.031042",
}

@ARTICLE{Smejkal2022c,
   author       = "L. Šmejkal  and A. Birk Hellenes and R. González-Hernández and J. Sinova and T. Jungwirth", 
   year         = "2022",
   journal      = "Phys. Rev. X",
   volume       = "12",
   pages        = "011028",
   title        = "Giant and tunneling magnetoresistance in unconventional collinear antiferromagnets with nonrelativistic spin-momentum coupling",
   doi = "https://doi.org/10.1103/PhysRevX.12.011028",
}

@ARTICLE{Amin2024,
   author       = "O. J. Amin and A. D. Din and E. Golias and Y. Niu and A. Zakharov and S. C. Fromage and C. J. B. Fields and S. L. Heywood and R. B. Cousins and F. Maccherozzi and others", 
   year         = "2024",
   journal      = "Nature",
   volume       = "636",
   pages        = "348",
   title        = "Nanoscale imaging and control of altermagnetism in {MnTe}",
   doi = "https://doi.org/10.1038/s41586-024-08234-x",
}

@ARTICLE{Yamamoto2025,
   author       = "R. Yamamoto and  L. A. Turnbull and M. Schmidt and J. C. C. Filho and H. J. Binger and M. D. P. Martínez and M. Weigand and S. Finizio and Y. Prots and G. M. Ferguson and others", 
   year         = "2025",
   journal      = "Phys. Rev. Applied",
   volume       = "24",
   pages        = "034037",
   title        = "Altermagnetic nanotextures revealed in bulk {MnTe}",
   doi = "https://doi.org/10.1103/dp7v-qszq",
}

@ARTICLE{Cheng2024,
   author       = "Q. Cheng and Y. Mao and Q.-F. Sun", 
   year         = "2024",
   journal      = "Phys. Rev. B",
   volume       = "110",
   pages        = "014518",
   title        = "Field-free Josephson diode effect in altermagnet/normal metal/altermagnet junctions",
   doi = "https://doi.org/10.1103/PhysRevB.110.014518",
}

@ARTICLE{Yi2025,
   author       = "X.-J. Yi and Yue Mao and X. Lu and Q.-F Sun", 
   year         = "2025",
   journal      = "Phys. Rev. B",
   volume       = "111",
   pages        = "035423",
   title        = "Spin splitting Nernst effect in altermagnets",
   doi = "https://doi.org/10.1103/PhysRevB.111.035423",
}

@ARTICLE{Zhou2025,
   author       = "Z. Zhou and X. Cheng and M. Hu and R. Chu and H. Bai and L. Han and J. Liu and F. Pan and C. Song", 
   year         = "2025",
   journal      = "Nature (London)",
   volume       = "638",
   pages        = "645",
   title        = "Manipulation of the altermagnetic order in {CrSb} via crystal symmetry",
   doi = "https://doi.org/10.1038/s41586-024-08436-3",
}

@ARTICLE{Stafford1996,
   author       = "C. A. Stafford and N. S. Wingreen", 
   year         = "1996",
   journal      = "Phys. Rev. Lett.",
   volume       = "76",
   pages        = "1916",
   title        = "Resonant photon-assisted tunneling through a double quantum dot: an electron pump from spatial rabi oscillations",
   doi = "https://doi.org/10.1103/PhysRevLett.76.1916",
}

@ARTICLE{Yeyati1997,
   author       = "A. L. Yeyati and J. C. Cuevas and A. López-Dávalos and A. Martín-Rodero", 
   year         = "1997",
   journal      = "Phys. Rev. B",
   volume       = "55",
   pages        = "R6137",
   title        = "Resonant tunneling through a small quantum dot coupled to superconducting leads",
   doi = "https://doi.org/10.1103/PhysRevB.55.R6137",
}

@ARTICLE{Sun2002,
   author       = "Q.-F. Sun and H. Guo", 
   year         = "2002",
   journal      = "Phys. Rev. B",
   volume       = "66",
   pages        = "155308",
   title        = "Double quantum dots: Kondo resonance induced by an interdot interaction",
   doi = "https://doi.org/10.1103/PhysRevB.66.155308",
}

@MISC{Chakraborty2007,
   title        = "{T. Chakraborty}, editor, Charge Migration in {DNA} ({Springer Berlin Heidelberg, Berlin, Heidelberg, 2007}).",
   doi = "https://doi.org/10.1007/978-3-540-72494-0",
}

@ARTICLE{Jiang2025,
   author       = "B. Jiang and M. Hu and J. Bai and Z. Song and C. Mu and G. Qu and W. Li and W. Zhu and H. Pi and Z. Wei and others", 
   year         = "2025",
   journal      = "Nat. Phys.",
   volume       = "21",
   pages        = "754",
   title        = "A metallic room-temperature d-wave altermagnet",
   doi = "https://doi.org/10.1038/s41567-025-02822-y",
}

@ARTICLE{ZhangX2025,
   author       = "X. Zhang and P. Jiang and L.-Y. Xu and L. Wang and L. Liu and H.-M. Huang and T. Cao and Y.-L. Li", 
   year         = "2025",
   journal      = "Nano Lett.",
   volume       = "25",
   pages        = "16547",
   title        = "Giant Spin Splitting and Anisotropic Spin Polarization in 2D Altermagnet {Cr2O}",
   doi = "https://doi.org/10.1021/acs.nanolett.5c04779",
}

@ARTICLE{Han2024,
   author       = "L. Han and X. Fu and R. Peng and X. Cheng and J. Dai and L. Liu and Y. Li and Y. Zhang and W. Zhu and H. Bai and others", 
   year         = "2024",
   journal      = "Sci. Adv.",
   volume       = "10",
   pages        = "eadn0479",
   title        = "Electrical 180° switching of Néel vector in spin-splitting antiferromagnet",
   doi = "10.1126/sciadv.adn0479",
}

@ARTICLE{Zhu2025,
   author       = "D. Zhu and J. Lu and Y. Jiang and Z. Zheng and D. Wang and C. Zhou and J. Zhou and S. Chen and Y. Gu and L. Liu and others", 
   year         = "2025",
   journal      = "Nano Lett.",
   volume       = "25",
   pages        = "11",
   title        = "Observation of anomalous {Hall} effect in collinear antiferromagnet {IrMn}",
   doi = "https://doi.org/10.1021/acs.nanolett.4c06271",
}

@ARTICLE{YZhang2025,
   author       = "Y. Zhang and H. Bai and J. Dai and L. Han and C. Chen and S. Liang and Y. Cao and Y. Zhang and Q. Wang and W. Zhu and others", 
   year         = "2025",
   journal      = "Nat. Commun.",
   volume       = "16",
   pages        = "5646",
   title        = "Electrical manipulation of spin splitting torque in altermagnetic {RuO2}",
   doi = "https://doi.org/10.1038/s41467-025-60891-2",
}

@ARTICLE{Li2025,
   author       = "Z. Li and Z. Zhang and Y. Chen and S. Hu and Y. Ji and Y. Yan and J. Du and Y. Li and L. He and X. Wang and others", 
   year         = "2025",
   journal      = "Adv. Mater.",
   volume       = "37",
   pages        = "2416712",
   title        = "Fully field-free spin-orbit torque switching induced by spin splitting effect in altermagnetic {RuO2}",
   doi = "https://doi.org/10.1002/adma.202416712",
}

@ARTICLE{Leiviska2024,
   author       = "M. Leiviskä and J. Rial and A. Bad'ura and R. Lopes Seeger and I. Kounta and S. Beckert and D. Kriegner and I. Joumard and E. Schmoranzerová and J. Sinova and others", 
   year         = "2024",
   journal      = "Phys. Rev. B",
   volume       = "109",
   pages        = "224430",
   title        = "Anisotropy of the anomalous {Hall} effect in thin films of the altermagnet candidate {Mn5Si3}",
   doi = "https://doi.org/10.1103/PhysRevB.109.224430",
}

@ARTICLE{Rial2024,
   author       = "J. Rial and M. Leiviskä and G. Skobjin and A. Bad'ura and G. Gaudin and F. Disdier and R. Schlitz and I. Kounta and S. Beckert and D. Kriegner and others", 
   year         = "2024",
   journal      = "Phys. Rev. B",
   volume       = "110",
   pages        = "L220411",
   title        = "Altermagnetic variants in thin films of {Mn5Si3}",
   DOI = "https://doi.org/10.1103/PhysRevB.110.L220411",
}

@ARTICLE{Xing2008,
   author       = "Y. Xing and Q. F. Sun and J. Wang", 
   year         = "2008",
   journal      = "Phys. Rev. B",
   volume       = "77",
   pages        = "115346",
   title        = "Influence of dephasing on the quantum {Hall} effect and the spin {Hall} effect",
   DOI = "https://doi.org/10.1103/PhysRevB.77.115346",
}

@ARTICLE{Morita2003,
   author       = "T. Morita and S. Kimura", 
   year         = "2003",
   journal      = "J. Am. Chem. Soc.",
   volume       = "125",
   pages        = "8732",
   title        = "Long-range electron transfer over 4 nm governed by an inelastic hopping mechanism in self-assembled monolayers of helical peptides",
   DOI = "https://doi.org/10.1021/ja034872n",
}

@ARTICLE{Skourtis2005,
   author       = "S. S. Skourtis and I. A. Balabin and T. Kawatsu and D. N. Beratan", 
   year         = "2005",
   journal      = "Proc. Natl. Acad. Sci. U.S.A.",
   volume       = "102",
   pages        = "3552",
   title        = "Protein dynamics and electron transfer: Electronic decoherence and non-{Condon} effects",
   doi = "https://doi.org/10.1073/pnas.0409047102",
}

@BOOK{Datta1995,
   author       = "S. Datta", 
   title        = "Electronic Transport in Mesoscopic Systems",
   publisher    = "Cambridge University Press",
   edition      = "First",
   year         = "1995",
   doi = "https://doi.org/10.1017/CBO9780511805776",
}

@ARTICLE{Lee1981,
   author       = "D. H. Lee and J. D. Joannopoulos", 
   year         = "1981",
   journal      = "Phys. Rev. B",
   volume       = "23",
   pages        = "4988",
   title        = "Simple scheme for surface-band calculations. I",
   doi = "https://doi.org/10.1103/PhysRevB.23.4988",
}

@ARTICLE{Mishra2020,
   author       = "S. Mishra and A. K. Mondal and S. Pal and T. K. Das and E. Z. B. Smolinsky and G. Siligardi and R. Naaman", 
   year         = "2020",
   journal      = "J. Phys. Chem. C",
   volume       = "124",
   pages        = "10776",
   title        = "Length-dependent electron spin polarization in oligopeptides and {DNA}",
   doi = "https://doi.org/10.1021/acs.jpcc.0c02291",
}

@ARTICLE{Evers2022,
   author       = "F. Evers and A. Aharony and N. Bar-Gill and O. Entin-Wohlman and P. Hedegård and O. Hod and P. Jelinek and G. Kamieniarz and M. Lemeshko and K. Michaeli and others", 
   year         = "2022",
   journal      = "Adv. Mater.",
   volume       = "34",
   pages        = "2106629",
   title        = "Theory of chirality induced spin selectivity: Progress and challenges",
   doi = "https://doi.org/10.1002/adma.202106629",
}

@ARTICLE{Steele2013,
   author       = "G. A. Steele and F. Pei and E. A. Laird and J. M. Jol and H. B. Meerwaldt and L. P. Kouwenhoven", 
   year         = "2013",
   journal      = "Nat. Commun.",
   volume       = "4",
   pages        = "1573",
   title        = "Large spin-orbit coupling in carbon nanotubes",
   doi = "https://doi.org/10.1038/ncomms2584",
}

@ARTICLE{Jhang2010,
   author       = "S. H. Jhang and M. Marganska and Y. Skourski and D. Preusche and B. Witkamp and M. Grifoni and H. van der Zant and J. Wosnitza and C. Strunk", 
   year         = "2010",
   journal      = "Phys. Rev. B",
   volume       = "82",
   pages        = "041404",
   title        = "Spin-orbit interaction in chiral carbon nanotubes probed in pulsed magnetic fields",
   doi = "https://doi.org/10.1103/PhysRevB.82.041404",
}

\end{document}